\documentclass[epj]{svjour}

\usepackage{amssymb}
\usepackage{amsmath}
\usepackage{epsfig}

\renewcommand{\vec}[1]{\mathbf{#1}}
\newcommand{\wti}[1]{\widetilde{#1}}
\newcommand{\til}[1]{\widetilde{#1}}

\newcommand{\vv}[1]{\mathbf{#1}}
\newcommand{\pp}{^{(+)}}

\newcommand{\mrm}[1]{\mathrm{#1}}
\newcommand{\ee}[1]{\mathrm{e}^{#1}}
\newcommand{\aveb}[1]{\bigl\langle #1 \bigr\rangle}

\begin{document}
\title{Virial expansion for charged colloids and electrolytes}

\author{Andr{\'e} G.\ Moreira\thanks{Present address: 
        Materials Research Laboratory, 
        University of California, Santa Barbara, CA 93106, USA} 
and Roland R.\ Netz\thanks{Present address: 
             Sektion Physik, Ludwig-Maximilians-Universit\"at,
             Theresienstr. 37, 80333 M\"unchen, Germany}}

\institute{Max-Planck-Institut f{\"u}r Kolloid- und Grenzfl{\"a}chenforschung,
D--14424 Potsdam, Germany}
\date{Received: date / Revised version: date}
\abstract{Using a field-theoretic approach, we derive
the first few coefficients of the exact low-density (``virial'') expansion 
of a binary mixture of positively and negatively
charged hard spheres (two-component hard-core plasma, TCPHC). 
Our calculations are nonperturbative with respect to the diameters
$d_+$ and $d_-$ and charge valences $q_+$ and $q_-$ of
positive and negative ions. Consequently, our closed-form 
expressions for the coefficients of the free energy  and
activity can be used to treat dilute salt solutions, where typically
$d_+ \sim d_-$ and $q_+ \sim q_-$, as well as colloidal suspensions,
where the difference in size and valence between macroions and
counterions can be very large. 
We show how to map the TCPHC on a one-component
hard-core plasma (OCPHC) in the 
colloidal limit of large size and valence ratio, in which case
the counterions effectively form a neutralizing background. A  sizable discrepancy 
with the standard  OCPHC with uniform, rigid background is detected, which 
can be traced back to the fact that the counterions cannot penetrate the 
colloids. For the case of electrolyte solutions, we show how to
obtain the cationic and anionic radii
as independent parameters from experimental data for the activity coefficient.
 % END OF PACS codes
} %end of abstract
\maketitle

\section{Introduction}
\label{chapterD:intro}

The two-component hard-core plasma (TCPHC) has been used
for a long time as an idealized model for electrolyte solutions.
In this model, also known as ``primitive mod\-el,''
the ions are spherical particles that interact with each other via 
the Coulomb potential and a hard-core potential, which avoids the collapse of
oppositely charged particles onto each other. 
In the general asymmetric case, the positive ions have a charge
$q_+ \, e$ (where $e$ is the elementary charge) and ionic diameter $d_+$,
while the negative ions have a charge $- q_- e$ and diameter $d_-$. 
The particles are immersed in a structureless solvent whose presence is
felt only through the value of the dielectric constant of the medium, 
and the system is (globally) electroneutral.

The TCPHC in this formulation is a gross simplification of real systems.
When two ions are at distances of the order of the size of the solvent molecules, 
the assumption of a continuum solvent breaks down, giving rise to so-called
solvation forces\cite{israelachvili:book}. 
Such effects also lead to non-local contributions to the dielectric constant.
Since the dielectric constant of ions or colloids typically differs from 
the surrounding aqueous medium, one also expects dispersion forces to act
between these particles.
All these effects,
which contribute to ``ion-specific'' effects\cite{ninham-yaminsky:97},
and are possibly more relevant than previously thought\cite{alfridsson-al:2000},
  are not accounted for
in the TCPHC, which treats the solvent---usually water---as a 
structureless medium within which the charged particles are embedded, 
neglecting the molecular arrangement that occurs around the ions.

Nevertheless, even with such simplifications, the TCP\-HC is far from 
being amenable to an exact treatment. A better understanding of this model
is a necessary step if one wishes to develop more realistic
approaches to charged systems. In this article, we turn our
attention to the low-density, or virial, expansion of the TCPHC. 
Since we have exact results for the thermodynamic properties at
low concentrations using field-theoretic methods, 
we can obtain useful information on dilute
systems with otherwise arbitrary hard-core radii and charge valences
(like colloidal suspensions).
In particular, we can test various approximations 
to treat the TCPHC, like the mapping of a colloidal solution
on an effective one-component plasma.

Due to the long-range character of the Coulomb potential,
it is not easy to obtain the thermodynamic behavior of the 
TCPHC through the usual methods of statistical mechanics.
For example, it can be shown\cite{mayer:50,friedman:book,mcquarrie:book} that the
straightforward application of
the cluster expansion to the TCPHC leads to divergent virial coefficients.
Mayer\cite{mayer:50} proposed a solution to this problem through an infinite 
resummation of the cluster diagrams, carried out such that
the divergent contributions to the virial expansion are canceled.
With this, he was able to obtain explicitly
the first term in the virial (or low-density) expansion that goes beyond the ideal
gas, which turns out to be the well-known Debye-H{\"u}ckel limiting 
law\cite{debye-hueckel:23}.
Haga\cite{haga:53} carried the expansion further and went up to order 
$5/2$ in the ionic density.
More or less at the same time, Edwards\cite{edwards:59} also obtained the virial
expansion of the TCPHC by mixing cluster expansion and field theory. 
In both cases only equally-sized ions were considered.

The  aforementioned  methods typically
depend on drawing, counting and recollecting the cluster diagrams which give finite 
contributions to the expansion up to the desired order in the density.
This can be quite a formidable task, and 
unfortunately it is easy to ``forget'' diagrams that are 
relevant to the series (see for instance the comment on pp.\ 222--223 of 
Ref.~\cite{friedman:book}).
Also, the generalization to ions with different
sizes is rather complicated\cite{friedman:book}.
Besides, the final results are typically not obtained in closed form, i.e., the 
final expressions depend on infinite sums that usually have to be evaluated numerically,
which reflects the infinite diagrammatic resummation.

We generalize here a novel field-theoretic technique\cite{roland-orland:tcp}, 
introduced for the symmetric TCPHC ($q_+ = q_-$ and $d_+=d_-$). We obtain
the exact low-density expansion of the \emph{asymmetric} 
(both in size and charge) TCPHC. 
This method does not use the cluster expansion (and resummation) and yields analytic, 
closed-form results. We go up to order
$5/2$ in the volume fraction of a system where the sizes and the charge valences of
positive and negative ions are unconstrained, that is, the results  we obtain can
be applied, without modifications, to both electrolyte solutions 
(where anions and cations have approximately the same size and valence)
and to colloidal suspensions (where the macro- and counterions
have sizes and valences that can be different by orders of magnitude).
In the colloidal limit,
we demonstrate how to map the TCPHC on an effective 
 one-component hard-core plasma (OCPHC) and obtain the corrections due to the
 exclusion of the background (i.e. the counterions) from the colloidal particles. 
For electrolytes, we obtain effective ionic sizes in solution,
or thermodynamic diameters, from experimental data for the mean activity.
The method developed here allows, in principle, to measure the diameters of
cations and anions as 
\emph{independent} variables, given that the range of experimental data 
extends to low enough densities such that the expansion used here 
is valid. It should be noted that the diameters obtained in this way are fundamentally 
different from the hydrodynamic diameters.

This article is organized as follows. In section~\ref{chapterD:method} we 
describe in detail
the steps that lead to the low-density expansion. Readers which are not interested 
in this
derivation can go directly to section~\ref{chapterD:applications}, 
where we apply the expressions obtained to two particular
problems, viz., (i) colloids, where one of the charged species is much smaller 
and much less charged
than the other one and (ii) the mean activity coefficient (which is related 
to the exponential of the chemical potential) of electrolyte solutions, which
is available from experiments and can be compared to our results.
Finally, section~\ref{chapterD:conclusions} contains some concluding remarks.

%%%%%%%%%%%%%%%%%%%%%%%%%%%%%%%%%%%%%%%%%%%%%%%%%%

\section{The method}
\label{chapterD:method}

We begin our calculation by assuming a system with $N_+$ positively charged 
particles with charge valence $q_+$ and diameter $d_+$, and
$N_-$ negatively charged particles with charge valence $q_-$ and
diameter $d_-$. Global electroneutrality of the system will
be imposed at a later stage of the calculation. The canonical 
partition function $\mathcal{Z}$ is given by
\begin{equation}
  \label{partition1}
  \mathcal{Z}=\frac{1}{N_+! N_-!}
  \int \prod\limits_{i=1}^{N_+} \frac{{\rm d}\vec{r}_i^{(+)}}{\lambda^3_{T,+}}
  \prod\limits_{j=1}^{N_-} \frac{{\rm d}\vec{r}_j^{(-)}}{\lambda^3_{T,-}}
  \, \exp \Bigl(- \frac{\mathcal{H}}{k_B T} \Bigr)
\end{equation}
where $\lambda_{T,+}$ and $\lambda_{T,-}$ are the thermal wavelengths,
and $\vv{r}\pp_i$ and $\vec{r}_i^{(-)}$ are the positions of positively 
and negatively charged particles.
The Hamiltonian $\mathcal{H}$ is given by
\begin{multline}
  \label{hamiltonian}
   \frac{\mathcal{H}}{k_B \, T} 
   = -E_{\mrm{self}} + \frac{1}{2} 
   \sum_{\alpha, \beta =+,-}
  \int {\rm d}\vec{r} {\rm d}\vec{r}' \, 
\hat{\rho}_{\alpha}(\vv{r}) \omega_{\alpha \beta}(\vv{r}-\vv{r}')
  \hat{\rho}_{\beta}(\vv{r}') \\
  + \frac{1}{2} \int {\rm d}\vec{r} {\rm d}\vec{r}' \,
  [\hat{\rho}_{+}(\vv{r})-\hat{\rho}_{-}(\vv{r})] v_c(\vv{r}-\vv{r}')
  [\hat{\rho}_{+}(\vv{r}')-\hat{\rho}_{-}(\vv{r}')]
\end{multline}
where the charge-density operators of the ions are defined as
\begin{equation}
  \begin{split}
    \label{density}
    \hat{\rho}_{+}(\vv{r}) &= q_+ \sum\limits_{i=1}^{N_+} \delta(\vv{r}-\vv{r}\pp_i) \\
    \hat{\rho}_{-}(\vv{r}) &= q_- \sum\limits_{i=1}^{N_-} \delta(\vv{r}-\vv{r}\pp_i),
  \end{split}
\end{equation}
and $\delta(\vv{r}-\vv{r}')$ is the Dirac delta function.
The indices $\alpha$ and $\beta$ in Eq.~(\ref{hamiltonian}) stand for $+$ and $-$, 
and the sum over $\alpha$ and $\beta$ in Eq.~(\ref{hamiltonian})
runs over all possible permutations (viz. $++$, $--$, $-+$ and $+-$), 
i.e., we consider a different
short-ranged potential $\omega_{\alpha \beta}$ for each combination.
The Coulomb potential is given by
 $v_c(\vv{r}) = \ell_B / r$, where
$\ell_B \equiv e^2 / ( 4 \,\pi \, \varepsilon \, k_B \, T )$ is
the Bjerrum length, defined as the distance at which the electrostatic 
energy between two elementary charges equals the thermal energy $k_B \, T$.
Finally, $E_{\mrm{self}}$ is the self-energy of the system
\begin{equation}
  \label{selfE}
  E_{\mrm{self}}= \frac{N_+ q_+^2 }{2} [\omega_{++}(0) + v_c(0)] +
  \frac{N_- q_-^2}{2} [\omega_{--}(0) + v_c(0)]
\end{equation}
which cancels the diagonal terms in Eq.~(\ref{hamiltonian}).

We proceed by applying the Hubbard-Stratono\-vich transformation,
the essence of which  is given by
\begin{multline}
  \label{HS}
  \ee{ -\frac{1}{2}
    \int {\rm d}\vec{r} {\rm d}\vec{r}' \, \hat{\rho}(\vv{r}) v(\vv{r}-\vv{r}') 
    \hat{\rho}(\vv{r}')} = \\
  \frac{\int {\cal D}\phi \, 
    \ee{-\frac{1}{2} \int {\rm d}\vec{r} {\rm d}\vec{r}' \,
      \phi(\vv{r}) v^{-1}(\vv{r}-\vv{r}') \phi(\vv{r}') - \imath \int {\rm d}\vec{r} \,
      \phi(\vv{r}) \hat{\rho}(\vv{r})}}{\int {\cal D}\phi \,
    \ee{ -\frac{1}{2} \int {\rm d}\vec{r} {\rm d}\vec{r}' \,
      \phi(\vv{r}) v^{-1}(\vv{r}-\vv{r}') \phi(\vv{r}') }}
\end{multline}
where $v(\vv{r})$ is some general potential and $\int {\cal D}\phi$ denotes a
path integral over the fluctuating field $\phi$. While this transformation
can be used without problems when $v(\vv{r})$ is the Coulomb potential,
for a short-ranged potential this can be more problematic:
for instance, a hard-core potential does not even have a 
well-defined inverse function. 
We will anyway take this formal step for the short-ranged potential, and, 
as we will see later, the way we handle the resulting expressions leads to
finite (and consistent) results, viz., the virial 
coefficients\cite{note1}.

Applying Eq.~(\ref{HS}) to Eq.~(\ref{partition1}) we obtain
the partition function in field-theoretic form
\begin{equation}
  \label{partition2}
  \mathcal{Z}= \int \frac{{\cal D}\psi_+ \, {\cal D}\psi_-}{\mathcal{Z}_{\psi}} 
  \frac{{\cal D}\phi}{\mathcal{Z}_{\phi}}
  \, \ee{-\bar{\mathcal{H}}_0} \, W_+ \, W_-
\end{equation}
with the action
\begin{multline}
  \label{H0}
  \bar{\mathcal{H}}_0 = \frac{1}{2} \sum_{\alpha, \beta =+,-}
    \int {\rm d}\vec{r} {\rm d}\vec{r}' \, \psi_{\alpha}(\vv{r})
    \omega^{-1}_{\alpha \beta}(\vv{r}-\vv{r}') \psi_{\beta}(\vv{r}') \\
    +\frac{1}{2} \int {\rm d}\vec{r} {\rm d}\vec{r}' \, \phi(\vv{r}) 
    v_c^{-1}(\vv{r}-\vv{r}') \phi(\vv{r}').
\end{multline}
The inverse potentials are formally defined as the solution of the equation
\begin{equation}
  \sum\limits_{\beta=+,-} \int {\rm d}\vec{r}' \, \omega_{\alpha \beta}(\vv{r}-\vv{r}')
  \omega^{-1}_{\beta \gamma}(\vv{r}'-\vv{r}'') =
  \delta_{\alpha \gamma} \, \delta(\vv{r}- \vv{r}'')
\end{equation}
($\delta_{\alpha \gamma}$ is the Kronecker delta) and
\begin{equation}
  \int {\rm d}\vec{r}' \, v_c(\vv{r}-\vv{r}') v^{-1}_c(\vv{r}'-\vv{r}'')=
  \delta(\vv{r}-\vv{r}'').
\end{equation}
For the Coulomb potential, $v_c^{-1}(\vv{r}) = - \nabla^2 \delta(\vv{r}) / 4 \pi \ell_B$.
We also define
\begin{equation}
  \label{W+}
  W_{\alpha} \! = \! \frac{1}{N_{\alpha}!} \Biggl[ 
  \ee{ q_{\alpha}^2 \bigl[\omega_{\alpha \alpha}(0)+v_c(0) \bigr]/2}
  \int \frac{{\rm d}\vec{r}}{\lambda^3_{T,\alpha}} \, 
\ee{-\imath q_{\alpha} \bigl[\psi_{\alpha}(\vv{r})+
    \alpha \, \phi(\vv{r}) \bigr]} \Biggr]^{N_{\alpha}}
\end{equation}
and the normalization factors
\begin{equation}
  \mathcal{Z}_{\psi}= \int {\cal D}\psi_+ \, {\cal D}\psi_-
  \, \ee{-\frac{1}{2} \sum\limits_{\alpha \beta}
    \int {\rm d}\vec{r} {\rm d}\vec{r}' \, \psi_{\alpha}(\vv{r})
    \omega^{-1}_{\alpha \beta}(\vv{r}-\vv{r}') \psi_{\beta}(\vv{r}')}
\end{equation}
and
\begin{equation}
  \label{Zphi}
  \mathcal{Z}_{\phi}= \int {\cal D}\phi  \, 
  \ee{-\frac{1}{2} \int {\rm d}\vec{r} {\rm d}\vec{r}' \, \phi(\vv{r}) 
    v_c^{-1}(\vv{r}-\vv{r}') \phi(\vv{r}')}.
\end{equation}

In order to make the calculations simpler we use the grand-canonical ensemble. 
This is achieved through the transformation
\begin{equation}
  \label{partition3}
  \mathcal{Q} = \sum_{N_+, N_- =0}^\infty \Lambda_+^{N_+} \Lambda_-^{N_-} \mathcal{Z},
\end{equation}
where $\Lambda_+$ and
$\Lambda_-$ are, respectively, the fugacities (exponential of the chemical potential)
of the positively and negatively charged particles. 
We perform the sum over $N_+$ and 
$N_-$ without constraints, i.e., without imposing the electroneutrality condition
\mbox{$q_+ N_+ = q_- N_-$}. Imposing this condition before going to the
grand-canonical ensemble makes the calculations much more difficult.
Later, electroneutrality will be imposed
order-by-order in the low-density expansion
in a consistent way, any infinities arising from the 
non-neutrality of the system will then be automatically canceled.

In its full form, the grand-canonical partition function $\mathcal{Q}$ reads
\begin{multline}
  \label{partition4}
  \mathcal{Q} = \int \frac{{\cal D}\psi_+ \, {\cal D}\psi_-}{\mathcal{Z}_{\psi}} 
  \frac{D\phi}{\mathcal{Z}_{\phi}}
  \, \exp \Biggl(
  \frac{\Lambda_+}{\lambda^3_{T,+}} \int {\rm d}\vec{r} \, h_+(\vv{r}) \ee{-i q_+ \phi(\vv{r})}  \\
  + \frac{\Lambda_-}{\lambda^3_{T,-}} \int {\rm d}\vec{r} \, h_-(\vv{r}) \ee{i q_- \phi(\vv{r})} 
   -\bar{\mathcal{H}}_0 \Biggr),
\end{multline}
where $\bar{\mathcal{H}}_0$ is given in Eq.~(\ref{H0}). We defined
the local non-linear operator
\begin{equation}
  \label{h}
  h_{\alpha}(\vv{r}) \equiv \exp \Bigl(\frac{q_{\alpha}^2 }{2} \bigl[
  \omega_{\alpha \alpha}(0)+ v_c(0) \bigr] - \imath
    q_{\alpha} \psi_{\alpha}(\vv{r}) \Bigr)
\end{equation}
(as before, $\alpha$ stands for both $+$ and $-$). At this point we rescale the 
fugacities such that $\lambda_+ = \Lambda_+/\lambda^3_{T,+}$
and $\lambda_- = \Lambda_-/\lambda^3_{T,-}$, i.e., the fugacities have
from now on dimensions of inverse volume.

Introducing the Debye-H{\"u}ckel propagator, 
\begin{equation}
  \label{vDH-1}
  v_{\mrm{DH}}^{-1}(\vv{r}-\vv{r}') = v_c^{-1}(\vv{r}-\vv{r}') + 
   I_2 \delta(\vv{r}-\vv{r}')
\end{equation}
with the ionic strength
\begin{equation}
I_2 = q_+^2 \lambda_+ + q_-^2 \lambda_-
\end{equation}
and after some algebraic manipulations, we finally obtain
the grand-canonical free energy density. It is defined through 
$g \equiv - \ln \bigl( \mathcal{Q} \bigr) / V$, and reads
\begin{multline}
  \label{free-energy1}
  g = - \lambda_+ - \lambda_- -
  \frac{1}{2} I_2 v_c(0)  -
  \frac{1}{V} \ln \Bigl( 
  \frac{\mathcal{Z}_{\mrm{DH}}}{\mathcal{Z}_{\phi}} \Bigr) \\ 
  - \frac{1}{V} \ln \Bigl\langle \ee{\lambda_+ \int {\rm d}\vec{r} \, Q_+(\vv{r}) +
    \lambda_- \int {\rm d}\vec{r} \, Q_-(\vv{r})} \Bigr\rangle,
\end{multline}
where $V$ is the volume of the system and the brackets 
$\langle \cdots \rangle$ denote averages over the 
fluctuating fields $\phi$ and $\psi_\alpha$ with the propagators 
$\omega^{-1}_{\alpha \beta}$ and $v^{-1}_{\mrm{DH}}$.
$\mathcal{Z}_{\mrm{DH}}$  is defined as
\begin{equation}
  \label{ZDH}
  \mathcal{Z}_{\mrm{DH}}= \int {\cal D}\phi  \, 
  \ee{-\frac{1}{2} \int {\rm d}\vec{r} {\rm d}\vec{r}' \, \phi(\vv{r}) 
    v_{\mrm{DH}}^{-1}(\vv{r}-\vv{r}') \phi(\vv{r}')}
\end{equation}
and the functions $Q_+(\vv{r})$ and $Q_-(\vv{r})$ are defined by
\begin{equation}
  \label{Q+}
  Q_{\alpha}(\vv{r}) = h_{\alpha}(\vv{r}) \ee{-\imath \alpha \, q_{\alpha} \phi(\vv{r})} - 1
  + \frac{1}{2} q_{\alpha}^2 \phi^2(\vv{r}) - \frac{1}{2} q_{\alpha}^2 v_c(0).
\end{equation}

In the preceding steps we obtained the exact expression for the grand-canonical
free energy density $g$, still without imposing electroneutrality. 
In order to obtain the low-density expansion of the free energy, we now
(i) expand $g$ in powers of $\lambda_+$ and $\lambda_-$ 
(up to a order $5/2$),
(ii) calculate the concentrations of positive and negative 
particles and impose
electroneutrality \emph{consistently}, order-by-order, and (iii)
make a Legendre transformation back to the canonical ensemble.

\subsection{Expansion in powers of the fugacities}

We start the expansion of $g$ by noting that the Fourier transform of
the Coulomb potential is $\wti{v}_c(\vv{k}) = 4 \pi \ell_B /k^2$. Using this, one
is able to express $v_c(0)$ as
\begin{equation}
  \label{vc(0)}
  v_c(0) = \int \frac{{\rm d} \vv{k}}{(2 \pi )^3} \frac{4 \pi \ell_B}{k^2}.
\end{equation}
It is easy to show that
\begin{equation}
  \label{logZZ}
  \frac{1}{V} \ln \Biggl( \frac{\mathcal{Z}_{\mrm{DH}}}{\mathcal{Z}_{\phi}} 
  \Biggr)  =
  -\frac{1}{2} \int \frac{{\rm d}\vv{k}}{(2 \pi )^3} \ln 
  \Biggl( 1 + \frac{4 \pi \ell_B I_2}{k^2} \Biggr).
\end{equation}
Using Eqs.~(\ref{vc(0)}) and (\ref{logZZ}), one finds
\begin{equation}
  \label{g_exp_1}
  \frac{1}{2} I_2 v_c(0) +
  \frac{1}{V} \ln \Biggl( \frac{\mathcal{Z}_{\mrm{DH}}}{\mathcal{Z}_{\phi}} \Biggr)
  = \frac{1}{12 \pi} \Bigl[ 4 \pi \ell_B I_2 \Bigr]^{3/2}.
\end{equation}
Introducing the dimensionless quantity 
\begin{equation}
{\Delta}v_0 = \sqrt{4 \pi \ell_B^3 I_2}
\end{equation} 
and expanding the last term in the rhs of
Eq.~(\ref{free-energy1}) in cumulants of $Q_+(\vv{r})$ and $Q_-(\vv{r})$, one obtains
the low-fugacity expansion
\begin{multline}
  \label{free-energy2}
  g = - \lambda_+ - \lambda_- - \frac{{\Delta}v_0^3}{12 \pi \ell_B^3} 
  -\lambda_+ Z^{+}_1 - \lambda_- Z^{-}_1 - \frac{\lambda_+^2}{2} Z_2^{++} \\
  - \frac{\lambda_-^2}{2} Z_2^{--} - \lambda_+ \lambda_- Z_2^{+-} +
  O(\lambda^3),
\end{multline}
where the coefficients are given by
\begin{equation}
  \label{z1+}
  Z_1^{+}=\frac{1}{V} \int {\rm d}\vec{r} \, \bigl\langle Q_+(\vv{r}) \bigr\rangle
\end{equation}
and  an analogous formula for $Z_1^{-}$, and 
\begin{equation}
  \label{z1++}
  Z_2^{++}=\frac{1}{V} \int {\rm d}\vec{r} {\rm d}\vec{r}' \,
  \Bigl\{ \aveb{Q_+(\vv{r}) Q_+(\vv{r}')} -
  \aveb{Q_+(\vv{r})} \aveb{Q_+(\vv{r}')} \Bigr\}
\end{equation}
and similar formulas for $Z_2^{--}$ and $Z_2^{+-}$.
The symbol $O(\lambda^3)$ in Eq.~(\ref{free-energy2}) means
that any other contribution to the expansion will be of
order 3 or higher, i.e., with terms like $\lambda_+^3$, 
$\lambda_+^2 \lambda_-$, etc.
Clearly, the expectation values in Eqs. (\ref{z1+}) and (\ref{z1++}) contain additional
dependencies on the fugacity $\lambda_\alpha$ via the DH propagator, Eq.(\ref{vDH-1}),
but, and this stands at the very core of our method, all expectation values can itself 
be expanded with respect to  $\lambda_\alpha$ and have  finite values as 
$\lambda_\alpha \rightarrow 0$

In order to do a full expansion of Eq.~(\ref{free-energy2}), 
one needs first to calculate the coefficients $Z_1^+$, etc., 
in Eq.~(\ref{free-energy2}),
for which  the averages given in the appendix are needed,
cf.\ Eq.~(\ref{a_av_1}--\ref{a_av_last}).
Since $v_c(0) - v_{\mrm{DH}}(0) = {\Delta}v_0$,  we obtain
\begin{equation}
  \label{Z1}
  Z_1^{\alpha} = \ee{q^2_{\alpha} {\Delta}v_0 / 2} - 1 - \frac{1}{2} q_{\alpha}^2 {\Delta}v_0
\end{equation}
and
\begin{multline}
  \label{Zlast}
  Z_2^{\alpha \beta} = \int {\rm d}\vec{r} \, \Biggl\{
  \ee{[q_{\alpha}^2 + q_{\beta}^2] {\Delta}v_0 / 2} 
  \Bigl[ \ee{-q_{\alpha} q_{\beta} [\omega_{\alpha \beta}(\vv{r}) + \alpha \beta \,
    v_{\mrm{DH}}(\vv{r})]} 
  -1 \Bigr] \\ + \frac{1}{2} q_{\alpha}^2 q_{\beta}^2 v_{\mrm{DH}}^2(\vv{r}) 
  \Bigl[ 1 - \ee{q_{\alpha}^2 {\Delta}v_0 /2} - \ee{q_{\beta}^2 {\Delta}v_0 /2} \Bigr] \Biggr\}.
\end{multline}
Note that
\begin{equation}
  \label{vDH}
  v_{\mrm{DH}}(\vv{r}) = \frac{\ell_B}{r} \ee{-{\Delta}v_0 r / \ell_B}
\end{equation}
was defined (through its inverse function) in Eq.~(\ref{vDH-1}). We now introduce the
hard-core through the short-range potentials
\begin{equation}
  \label{hard-core}
  \omega_{\alpha \beta}(\vv{r}) = 
  \begin{cases}
    + \infty& \text{if } r < \bigl( d_{\alpha} + d_{\beta} \bigr) / 2, \\
    0 & \text{otherwise,}
  \end{cases}
\end{equation}
where the indices $\alpha$ and $\beta$ stand again for $+$ and $-$;
$d_+$ and $d_-$ are respectively the (effective) ionic diameters
of the positive and negative particles.

Since
the expressions for $Z_1^+$, etc., do depend on $\lambda_+$ and
$\lambda_-$, and 
in order to have a consistent expansion of $g$ in the fugacities,
one has to expand Eqs.~(\ref{Z1}--\ref{Zlast}) in powers of $\lambda_+$ and $\lambda_-$,
up to the appropriate order, before inserting them into $g$, 
Eq.~(\ref{free-energy2}). By doing
this consistently up to order $5/2$ in the fugacities, one obtains
the rescaled grand-canonical free energy density
\begin{multline}
  \label{free-energy3}
    \wti{g} \equiv d_+^3 g = \\ 
    -\wti{\lambda}_+ - \wti{\lambda}_- 
    - m_1 \wti{\lambda}_+^2 - m_2 \wti{\lambda}_-^2 -
    m_3 \wti{\lambda}_+ \wti{\lambda}_- 
    - \Bigl[ n_1 \wti{\lambda}_+^2 + \\
    n_2 \wti{\lambda}_-^2 + n_3 \wti{\lambda}_+ 
    \wti{\lambda}_- \Bigr] {\Delta}v_0  -
    \Bigl[ p_1 \wti{\lambda}_+^2 + 
    p_2 \wti{\lambda}_-^2  + p_3 \wti{\lambda}_+ 
    \wti{\lambda}_- \Bigr] \ln {\Delta}v_0 \\ 
    -\Bigl[ r_1 \wti{\lambda}_+^2  + 
    r_2 \wti{\lambda}_-^2  + r_3 \wti{\lambda}_+ \wti{\lambda}_- \Bigr] 
    {\Delta}v_0 \ln {\Delta}v_0 
    - \Bigl[ s_1 \wti{\lambda}_+ \\ + s_2 \wti{\lambda}_- \Bigr] {\Delta}v_0^2  -
    \Bigl[ t_0 + t_1 \wti{\lambda}_+ + t_2 \wti{\lambda}_- \Bigr] {\Delta}v_0^3 
    + \Omega_0 \Bigl[ q_+ \wti{\lambda}_+ - q_- \wti{\lambda}_- \Bigr]  \\ -
    {\Delta}v_0 \Biggl\{ \Omega_1 \Bigl[ q_+ \wti{\lambda}_+ - q_- 
    \wti{\lambda}_- \Bigr]^2 -
    \Omega_0 \Bigl[ q_+^4 \wti{\lambda}_+^2 + q_-^4 \wti{\lambda}_-^2 \\ - 
    2 q_+ q_- \Bigl[ \frac{q_+^2+q_-^2}{2} \Bigr] \wti{\lambda}_+ 
    \wti{\lambda}_- \Bigr]
    \Biggr\}+ O(\tilde{\lambda}^3),
\end{multline}
where we used dimensionless fugacities
\begin{equation}
  \wti{\lambda}_+ \equiv d_+^3 \lambda_+ 
\end{equation}
and 
\begin{equation}
  \wti{\lambda}_- \equiv d_+^3 \lambda_-.
\end{equation}
In this expansion, we utilized that ${\Delta}v_0$ scales
like $\wti{\lambda}^{1/2}$. The coefficients $m_1$, etc. are given explicitly
in the appendix, Eqs.~(\ref{a_coeff_1}--\ref{a_coeff_last}). 

The coefficients $\Omega_0$ and $\Omega_1$ are given by the
divergent integrals
\begin{equation}
  \begin{split}
    \Omega_0 &= 2 \pi \ell_B \int\limits_0^{\infty} {\rm d}r \, r  \\
    \Omega_1 &= 2 \pi \int\limits_0^{\infty} {\rm d}r \, r^2.
  \end{split}
\end{equation}
These terms are present in Eq.~(\ref{free-energy3})
because global charge neutrality has not been yet demanded. By imposing
this condition, these divergent terms cancel exactly, as is
shown next.

\subsection{Imposing electroneutrality}
\label{chapterD:neutral}

The electroneutrality condition ensures that the global charge
of the system is zero, i.e., $q_+ N_+ = q_- N_-$.
In the grand-canonical ensemble, $N_+$ and $N_-$ are no longer
fixed numbers but average values. This means that the electroneutrality condition
in the grand-canonical ensemble is given by
\begin{equation}
  \label{neutro}
  q_+ \langle N_+ \rangle = q_- \langle N_- \rangle.
\end{equation}
Defining the rescaled ion density
\begin{equation}
\wti{c}_+ = d_+^3 \langle N_+ \rangle / V,
\end{equation}
it is easy to show that
\begin{equation}
  \label{concentra}
  \wti{c}_+ = - \wti{\lambda}_+ \frac{\partial \wti{g}}
  {\partial \wti{\lambda}_+}
\end{equation}
with an analogous formula for $\wti{c}_-$; $\wti{c}_+$ is
the volume fractions of the positive ions\cite{note_volfrac}.
As one imposes Eq.~(\ref{neutro}), the fugacities
will depend on each other in a non-trivial way such that the system is,
on average, neutral. If the system were totally symmetric (i.e., $q_+=q_-$
and $d_+=d_-$), this dependence would be given by the relation
$\wti{\lambda}_+ = \wti{\lambda}_-$\cite{roland-orland:tcp}. However, 
this is not the case here: one
needs to find the relation between the fugacities order-by-order.

First, assume that $\wti{\lambda}_-$ can be expanded in terms of
$\wti{\lambda}_+$ such that
\begin{multline}
  \label{lambda-}
  \wti{\lambda}_- = a_0 \wti{\lambda}_+ 
  + a_1 \wti{\lambda}_+^{3/2} + 
  a_2 \wti{\lambda}_+^2 +
  a_3 \wti{\lambda}_+^2 \ln \wti{\lambda}_+ \\ 
  + a_4 \wti{\lambda}_+^{5/2} +
  a_5 \wti{\lambda}_+^{5/2} \ln \wti{\lambda}_+ + 
  O(\wti{\lambda}_+^3).
\end{multline}
This is inspired by the expanded form of the grand-canonical free
energy $\wti{g}$, Eq.(\ref{free-energy3}).

After calculating $\wti{c}_+$ and $\wti{c}_-$ from $\wti{g}$,
Eq.~(\ref{free-energy3}), according to Eq.(\ref{concentra}), and
insertion into the electroneutrality condition
Eq.~(\ref{neutro}), we substitute the fugacity  $\wti{\lambda}_-$
by its expanded form Eq.~(\ref{lambda-}).
This leads to the expanded form (up to order $\wti{\lambda}_+^{5/2} 
\ln \wti{\lambda}_+$) of
the electroneutrality condition. Solving it consistently, order-by-order,
yields the values of  the coefficients in Eq.(\ref{lambda-}), which ensure
electroneutrality order by order.  For 
instance, at linear order in $\wti{\lambda}_+$, 
the expanded form of Eq.~(\ref{neutro}) reads
\begin{equation}
  q_+ - a_0 q_- = 0,
\end{equation}
which naturally gives $a_0 = q_+ / q_-$. With the knowledge of $a_0$, one can
solve the next-order term (in this case $\wti{\lambda}_+^{3/2}$) and obtain the 
value of $a_1$, and so on. The resulting coefficients $a_0$ up to $a_5$ are given
in the appendix---cf.\ Eqs.~(\ref{a_a0}--\ref{a_a5}).

As this order-by-order neutrality condition is imposed,
one notices that the terms
$\Omega_0$ and $\Omega_1$ in Eq.~(\ref{free-energy3}) are exactly canceled
in a natural way, without any further assumptions.
The resulting expression for $\wti{g}$, now expanded
only in one of the fugacities (in this case $\wti{\lambda}_+$), is then a well
behaved expansion (we omit this expression here since it is quite lengthy).
With this, one can finally obtain the canonical free energy (as a density expansion)
through a Legendre transform.

\subsection{The canonical free energy}

In order to obtain the free energy in the canonical ensemble, 
we use the back-Legendre-transformation
\begin{equation}
  \label{legendre}
  \wti{f} = \wti{g} + \wti{c}_+ 
  \ln \bigl( \wti{\lambda}_+ \bigr) + 
  \wti{c}_- \ln \bigl( \wti{\lambda}_-  \bigr),
\end{equation}
where  the dimensionless canonical free energy density is
\begin{equation}
  \wti{f} \equiv d_+^3 F / V k_B T 
\end{equation}
(F is the canonical free energy). Note
that at this point $\wti{\lambda}_-$ is a function of $\wti{\lambda}_+$, according
to Eq.~(\ref{lambda-}). 

\begin{figure*}
  \begin{center}
    \epsfig{file=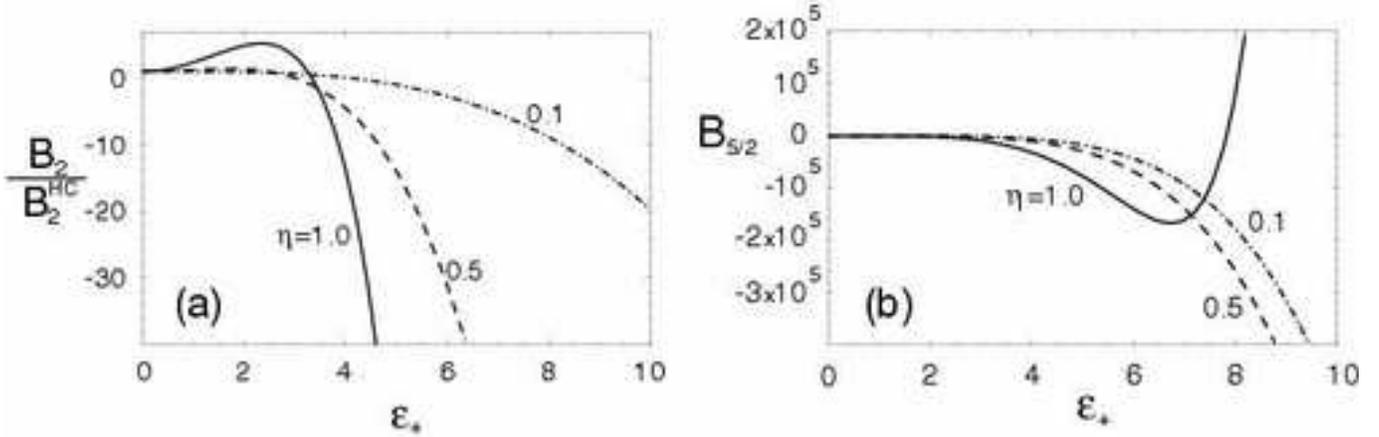, width=\textwidth}
  \end{center}
  \caption{Coefficients (a) $B_2$ and (b) $B_{5/2}$ of the
free energy expansion Eq.~(\ref{f}) as functions of $\epsilon_+$ 
for different values of $\eta \equiv q_-/q_+$ 
(with $d_+ = d_-$ or $\xi=1$). 
Notice that $B_2$ is normalized by
the second virial coefficient $B_2^{\rm HC}$
of a pure two-component hard-core gas.}
  \label{fig:fig1}
\end{figure*}

The first step to obtain $\wti{f}$ is to invert the
expression given in Eq.~(\ref{concentra}) such that $\wti{\lambda}_+$ is
obtained as an expansion in $\wti{c}_+$, 
neglecting any terms of order $\wti{c}_+^3$ 
or higher. With this, we obtain 
$\wti{\lambda}_+ = \wti{\lambda}_+(\wti{c}_+)$: 
plugging this into Eq.(\ref{legendre}),
we finally obtain $\wti{f}$, which reads
\begin{multline}
  \label{f}
  \wti{f}=\wti{f}_{\mrm{id}} + B_{\mrm{DH}} \wti{c}_+^{3/2} +
  B_2 \wti{c}_+^2 + B_{2\mrm{log}} \wti{c}_+^2 \ln \wti{c}_+
  \\ + B_{5/2} \wti{c}_+^{5/2} + B_{5/2\mrm{log}} \wti{c}^{5/2}_+ \ln \wti{c}_+
  +O\bigl(\tilde{c}_+^3 \bigr).
\end{multline}
Defining  the valence ratio parameter
\begin{equation}
  \eta \equiv q_- / q_+,  
\end{equation}
the diameter ratio parameter
\begin{equation}
  \xi \equiv d_- / d_+,
\end{equation}
and the coupling parameter 
\begin{equation}
  \epsilon_+ \equiv q_+^2 \ell_B / d_+  
\end{equation}
(which is the ratio between the Coulomb energy at contact between two
positive ions and the thermal energy $k_B T$), the coefficients in
Eq.~(\ref{f}) can be explicitly written as
\begin{equation}
  \label{ideal}
  \wti{f}_{\mrm{id}} = \wti{c}_+ \ln \wti{c}_+ + \frac{\wti{c}_+}{\eta} 
  \ln \Bigl( \frac{\wti{c}_+}{\eta} \Bigr) - 
  \Bigl[ 1 + \frac{1}{\eta} \Bigr] \wti{c}_+
\end{equation}
which is the ideal contribution to the free energy,
\begin{equation} \label{BDH}
  B_{\mrm{DH}}= - \frac{2}{3} 
  \sqrt{\pi  \epsilon_+^3 [1+\eta]^3}
\end{equation}
which is the coefficient of the Debye-H{\"u}ckel limiting law term (order $3/2$ in 
$\til{c}_+$). It is useful to define the function
\begin{equation} \label{Hdefine}
  H(x) = \frac{11}{6} - 2 \gamma + \frac{1}{x^3} \ee{-x} 
  \bigl[ 2 - x + x^2 \bigr] - \Gamma(0,x) - \ln x,
\end{equation}
where $\gamma$ is the Euler's constant and $\Gamma(a,b)$ is the
incomplete Gamma-function\cite{abramowitz:book}. Using $H(x)$,
the higher order coefficients can be explicitly written as
\begin{multline} \label{B2}
  B_2 = - \frac{\pi}{3} \epsilon_+^3 \Biggl\{ - H(\epsilon_+) - \ln \epsilon_+
  + 2 \eta^2 \Biggl[ H\Bigl(-\frac{2 \eta \epsilon_+}{1+\xi} \Bigr)  \\ + 
  \ln\Bigl( \frac{2 \eta \epsilon_+}{1+\xi} \Bigr) \Biggr]
  - \eta^4 \Biggl[ H \Bigl( \frac{\eta^2 \epsilon_+}{\xi} \Bigr) +
  \ln\Bigl( \frac{\eta^2 \epsilon_+}{\xi} \Bigr) \Biggr] \\
  - 2 \eta^2 \bigl[ 1 - \eta^2 \bigr] \ln\eta 
  + \frac{1}{2} \bigl[ 1-\eta^2 \bigr]^2
  \ln \bigl( 36 \pi \epsilon_+^3 [1+\eta] \bigr) \Biggl\}
\end{multline}
\begin{equation} \label{B2log}
  B_{2\mrm{log}} = - \frac{\pi}{6} \epsilon_+^3 \bigl[ 1-\eta^2 \bigr]^2,
\end{equation}
\begin{multline}
  B_{5/2} = \frac{2}{3} \bigl[ \pi \epsilon_+^3 \bigr]^{3/2}
  \bigl[1 + \eta \bigr]^{1/2} \Biggl\{
  \frac{5}{8} + H\bigl(\epsilon_+ \bigr) + \ln \bigl( \epsilon_+ \bigr) \\
  + \eta^6 \Biggl[ \frac{5}{8} +
  H\Bigl( \frac{\eta^2 \epsilon_+}{\xi} \Bigr) +
  \ln \Bigl( \frac{\eta^2 \epsilon_+}{\xi} \Bigr) \Biggr]+
  2 \eta^3 \Biggl[ \frac{5}{8} + H\Bigl(-\frac{2 \eta \epsilon_+}{1+\xi} \Bigr) \\
  + \ln\Bigl(\frac{2 \eta \epsilon_+}{1+\xi} \Bigr) \Biggr]
  + \frac{1}{8} \bigl[ 1 + \eta \bigr]^2  \bigl[ 5 - 12 \eta +
  17 \eta^2 - 12 \eta^3 + 5 \eta^4 \bigr] \\
  - 2 \eta^3 \bigl[ 1 + \eta^3 \bigr] \ln \eta 
  - \frac{1}{2} \bigl[ 1 + \eta^3 \bigr]^2 
  \ln\Bigl( 64 \pi \epsilon_+^3 \bigl[ 1 + \eta \bigr] \Bigr) 
  \Biggr\}
\end{multline}
and 
\begin{equation}
  B_{5/2\mrm{log}} = - \frac{1}{3} \bigl[ \pi \epsilon_+^3 \bigr]^{3/2}
  \bigl[1 + \eta \bigr]^{1/2} \bigl[ 1 + \eta^3 \bigr]^2.
\end{equation}

The free energy Eq.~(\ref{f}) is the exact low density expansion of the
asymmetric TCPHC and constitutes the main result of this paper. 
The only parameter that is demanded to be small is  the ion density
$\wti{c}_+$, that means, this result is non-perturbative in the coupling
$\epsilon_+$, charge ratio $\eta$ and  size ratio $\xi$.

We chose the positive ions as the ``reference species''
(i.e., the expansion is done with respect to $\wti{c}_+$) without any loss
of generality, since the relation between
$\wti{c}_+$ and $\wti{c}_-$ is fixed through the electroneutrality condition.
As consistency checks, we notice that our expression 
for $\wti{f}$ is symmetric, as expected, with respect to the
simultaneous exchange of $d_+$ with $d_-$ and $q_+$ with $q_-$.
Also, in the limit $d_+ = d_-$ and $q_+ = q_-$, we recover the same expression
as previously calculated in Ref.~\cite{roland-orland:tcp} for totally symmetric systems.
Finally, as one turns off the charges in the system (or equivalently, 
as one takes the limit $\epsilon_+ \rightarrow 0$), 
the pure hard-core fluid is recovered, i.e., 
$\wti{f}$ becomes the usual virial expansion with $B_{3/2}$ and
$B_{5/2}$ equal to zero and $B_2$ given by the second virial coefficient 
of a two-component hard-core gas, 
$B_2^{\rm HC} = B_2(\epsilon_+ \rightarrow 0)$,
which in our units reads
\begin{equation} \label{BHC}
  B_2^{\mrm{HC}}=\frac{\pi}{3} 
  \biggl[ 2 + \frac{2 \xi^3}{\eta^2} + \frac{[1+\xi]^3}{2 \eta} \biggr].
\end{equation}
This limit can be also understood
as the high-temperature regime: as the thermal energy largely exceeds
the Coulomb energy at contact, the hard core interaction becomes the 
only relevant interaction between the particles.

%%%%%%%%%%%%%%%%%%%%%%%%%%%%%%%%%%%%%%%%%%%%%%%%%%%%%%%%%%%
\section{Results}
\label{chapterD:applications}

\begin{figure*}
  \begin{center}
    \epsfig{file=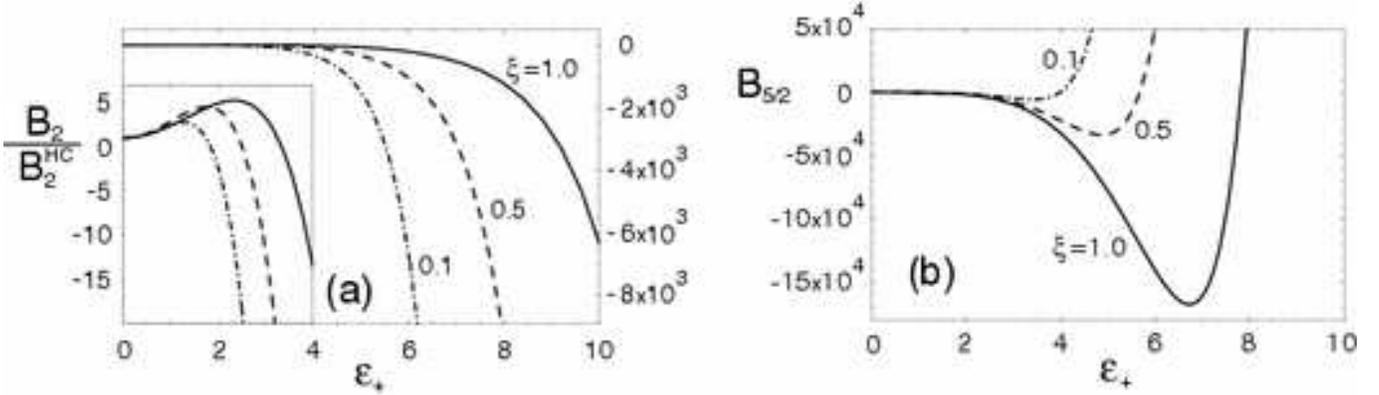, width=\textwidth}
  \end{center}
  \caption{Coefficients (a) $B_2$ and (b) $B_{5/2}$ of the
    free energy Eq.~(\ref{f}) as functions of $\epsilon_+$ 
    for different values of $\xi \equiv d_-/d_+$ 
    (with $q_+ = q_-$). Notice that $B_2$ is normalized by
    the second virial coefficient $B_2^{\rm HC}$
 of a pure two-component hard-core gas. The inset
    in (a) shows the behavior of $B_2$ close to $\epsilon_+ =0$
    on a different scale.}
  \label{fig:fig2}
\end{figure*}

\subsection{The virial coefficients}

The behavior of the coefficients $B_2$ and $B_{5/2}$ as functions of
the coupling parameter $\epsilon_+$ are depicted in 
Figs.~\ref{fig:fig1} and \ref{fig:fig2},
the behavior of $B_{\rm 2log}$ and $B_{\rm 5/2log}$ is rather trivial and 
not shown graphically.
 In Fig.~\ref{fig:fig1}
the ionic diameters of positive and negative ions
are kept equal, $d_+=d_-$, and the ratio between the charge valences, $\eta$, is
varied, while in Fig.~\ref{fig:fig2} the charge valences are equal 
and the ratio between the ionic
diameters is varied. These figures highlight the fact that
both coefficients diverge as $\epsilon_+$ goes to infinity.
In this limit,
\begin{equation}
  \label{B2_asy}
  B_2 \approx - \frac{\pi}{4} \frac{[1 + \xi]^4}{\eta^2} 
  \frac{1}{\epsilon_+} \exp\Bigl(\frac{2 \eta  \epsilon_+}{1+\xi} \Bigr)
\end{equation}
and
\begin{equation}
  \label{B52_asy}
  B_{5/2} \approx \frac{\pi^{3/2}}{2} \frac{[1+\xi]^4 \sqrt{1+\eta}}{\eta}
  \sqrt{\epsilon_+} \exp\Bigl(\frac{2 \eta  \epsilon_+}{1+\xi} \Bigr).
\end{equation}
Note that $B_{5/2}$ diverges faster (and with opposite sign) 
than $B_2$, reflecting that this low-density expansion is badly converging,
as we will discuss further below.
The exponentially
divergent behavior of $B_2$ and $B_{5/2}$ when $\epsilon_+ \rightarrow \infty$ 
is due to the increasing importance of the interaction between
oppositely charged particles (ionic pairing)
as the coupling parameter increases\cite{bjerrum:26,fisher-levin:93,yeh-zhou-stell:96},
corresponding for instance to lower temperatures.
This is confirmed by noting that the argument in the exponential
occurring in both asymptotic forms Eqs.~(\ref{B2_asy}) and (\ref{B52_asy}),
viz.\ $2 \, \eta \, \epsilon_+ / [1 + \xi]$, can be re-expressed
as $2 \, q_+ \, q_- \, \ell_B / [d_+ + d_-]$, which is
the coupling between positive and negative ions 
(in this case, the ratio between the Coulomb contact 
energy between oppositely charged ions and the thermal energy $k_B T$).

\begin{figure*}
  \begin{center}
    \epsfig{file=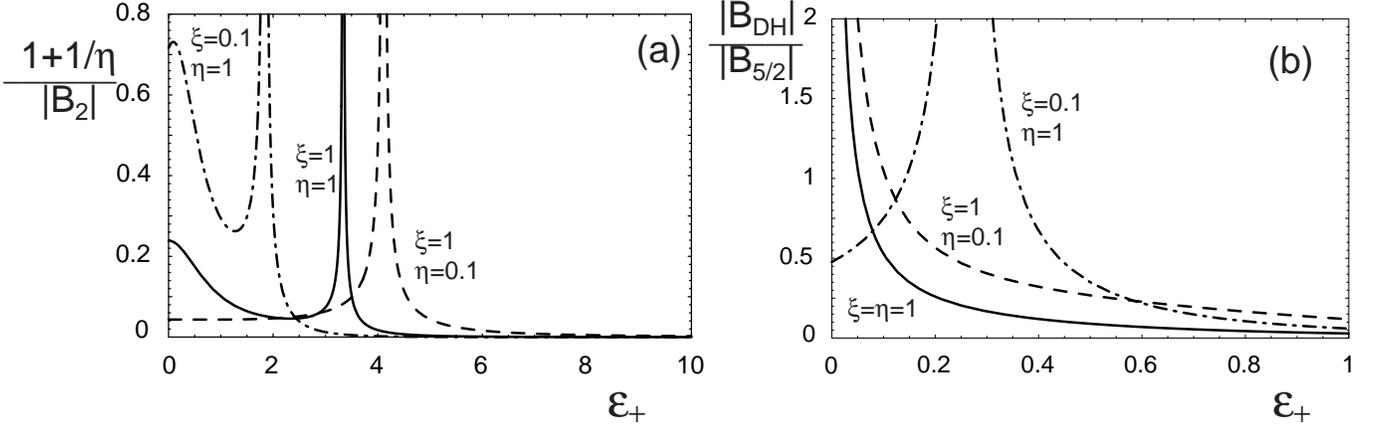, width=\textwidth}
  \end{center}
  \caption{Ratios of the coefficents in the low-density 
expansion, which serve as rough estimates of the maximal rescaled
ion density $\tilde{c}_+$ up to which the expansion is valid
(see text). }
  \label{figratio}
\end{figure*}

To estimate roughly the ionic density  $\tilde{c}_+$ up to which the 
expansion in Eq.~(\ref{f}) is expected to be valid, we use the following
simple criterion: the terms proportional to $B_{\rm DH}$ and
$B_{5/2}$ give the same contribution to the free energy when 
$|B_{\rm DH} |\tilde{c}_+^{3/2} =| B_{5/2} | \tilde{c}_+^{5/2}$
or, in other words, at a critical density $\tilde{c}_+ =
|B_{\rm DH}/B_{5/2}|$. Note that we set up this criterion separately
for the integer and fractional coefficients, since the scaling
behavior for the two different classes is very different and no
meaningful results can be obtained by mixing them. The analogous
criterion for the integer terms leads to a critical density 
$\tilde{c}_+ = (1+1/\eta) /|B_{2}|$ where for the ideal contribution in
Eq. \ref{ideal} we replace the logarithmic term by a linear one
(which is the ideal contribution to the 
osmotic pressure). We tacitly assume that the higher-order terms (which
we have not calculated) show the same relative behavior, an assumption
which seems plausible to us but which we cannot check.
 In Fig. \ref{figratio} we show the estimate for
the critical densities obtained from
 a) the ratio of the integer terms and from
 b) the ratio of the fractional terms.
In a) the critical density settles at a finite value for vanishing
coupling parameter $\varepsilon_+ \rightarrow 0$, and decreases to 
zero for increasing coupling parameter (the divergence in the curves
at finite value of $\varepsilon_+$ is not significant since it is caused
by a change in sign of $B_2$.). This means that for small coupling
constants the integer coefficients are expected to show regular convergence
behavior for not too large concentrations. For large values of
$\tilde{c}_+$ on the other hand, the range of densities which 
can be described by the expansion quickly goes to zero.
In b) the displayed behavior is more complex.
The expansion of the coefficient $B_{5/2}$ for small values of
the coupling parameter $\varepsilon_+$ leads to
\begin{eqnarray}
B_{5/2}&=&  
\pi^{3/2} \sqrt{1+\eta} (\xi-1)^2(\xi+1) \varepsilon_+^{3/2}  \\
&&-\pi^{3/2} \sqrt{1+\eta} (2+2\eta^2 \xi^2+\eta(1+\xi)^2) \nonumber
\varepsilon_+^{5/2}
\end{eqnarray}
plus terms with scale as higher powers in $\varepsilon_+$.
The important point is that except for the singular case
$\xi=1$, that is for particles of identical radius, the leading
term of $B_{5/2}$ scales as $ \varepsilon_+^{3/2}$ and thus identical
to the lower leading term $B_{\rm DH}$.
In the case $\xi=1$ the leading term scales as $\varepsilon_+^{5/2}$
and thus shows a different scaling behavior than  $B_{\rm DH}$.
In Fig. \ref{figratio}b) this difference is cleary seen: 
For $\xi=1$ the curves diverge as $\varepsilon_+ \rightarrow 0$
and the convergence of the fractional density terms in the series is 
expected to be guaranteed. For $\xi =0.1$ the behavior is similar
to the results in a), showing a saturation at a finite value 
as $\varepsilon_+ \rightarrow 0$ (and an unsignificant divergence
at finite $\varepsilon_+$). All curves go to zero as 
the coupling parameter grows. In conclusion, the reliability 
of the low-density expansion becomes worse as the coupling parameter
increases, but one can always find a window of small densities 
within which the expansion should work.

Finally, we mention that in a different calculation scheme, based 
on an integral-equation-procedure within the so-called 
mean-spherical-model (MSM) approximation, similar results for the free
energy and other thermodynamic functions have been 
obtained\cite{waisman-lebowitz:72,Palmer,Blum}. Those results
also reproduce the limiting laws, namely the hard-core behavior
as the coupling parameter goes to zero, and the leading Debye-H\"uckel 
correction at low density, and give a quite satisfactory
description even for much larger values of the density. It is important
to note, however, that the next-leading (beyond Debye-H\"uckel)
terms in the low-density
expansion are not correctly reproduced by the MSM approximation.

\subsection{The colloidal limit}

In colloidal suspensions, flocculation or coagulation
(driven by attractive van der Waals interaction between 
colloidal particles) can be  prevented by the
presence of repulsive electrostatic forces. These 
suspensions are typically quite dilute, with volume fractions of 
colloidal particles usually not higher than a few percent. The macro-particles
normally have dimensions\cite{ottewill:review,gonnet-al:94,kurth-al:2000}
 ranging from 10 to 1000~nm and
charges of several thousands $e$ (elementary charge), with much smaller counterions that
have a charge of a few $e$. In such systems the charge
and size asymmetry between ions and counterions is immense; the
TCPHC with unconstrained charge and size asymmetry is a suitable model for
dilute colloidal suspensions, and the free energy Eq.~(\ref{f}) can be used to
study the thermodynamic behavior of such systems.

With this in mind, let us take the following limit: 
assume the parameters describing the positive ions (representing the colloids)
$d_+$, $q_+$ and $\wti{c}_+$ fixed, and make
both $\eta \equiv q_-/q_+$ and $\xi \equiv d_-/d_+$ vanishingly
small (this has to be done with some care, 
since the limit $\eta=0$ is not well-defined). 
One can then rewrite the free energy Eq.~(\ref{f}) up to second order in $\wti{c}_+$ as
\begin{multline}
  \label{f-collo}
  \wti{f}^{\mrm{cl}} = \wti{f}_{\mrm{id}} + 
  + B^{\mrm{cl}}_{\mrm{DH}} \wti{c}_+^{3/2} +
  ( B_2^{\mrm{HC,cl}} + B^{\mrm{cl}}_2) \wti{c}_+^2 \\
  + B^{\mrm{cl}}_{2\mrm{log}} \wti{c}_+^2 \log \wti{c}_+ +  O(\wti{c}_+^{5/2})
\end{multline}
where $\wti{f}_{\mrm{id}}$ is the ideal term Eq.~(\ref{ideal}) and 
\begin{equation}
  \label{f-collo-hc}
  B_2^{\mrm{HC,cl}} = \frac{\pi}{3} \biggl[ \frac{2 \xi^3}{\eta^2} + 
  \frac{[1+\xi]^3}{2 \eta} \biggr] 
\end{equation}
is the  hard core contribution due to the
counterion--counterion and macroion--counterion
interactions only, without the macroion--macroion contributions,
which explains  the difference to the full hard-core virial coefficient
in Eq.(\ref{BHC}). It is necessary to treat this term separately, as it does
not have a well-defined behavior in the double limit $\eta \rightarrow 0$
and $\xi \rightarrow 0$.
The other coefficients in Eq.(\ref{f-collo}) follow from 
Eqs.(\ref{BDH}), (\ref{B2}), (\ref{B2log}) by performing the limits
$B^{\mrm{cl}}_{\mrm{DH}} =  B_{\mrm{DH}} (\eta \rightarrow 0, \xi \rightarrow 0)$, 
$B^{\mrm{cl}}_{\mrm{2}} =  B_{\mrm{2}} (\eta \rightarrow 0, \xi \rightarrow 0)
-B_2^{\mrm{HC,cl}}$, 
$B^{\mrm{cl}}_{\mrm{2log}} =  B_{\mrm{2log}} (\eta \rightarrow 0, \xi \rightarrow 0)$, 
and are given by
\begin{equation}
  \label{BDHcollo}
  B^{\mrm{cl}}_{\mrm{DH}} = -\frac{2}{3} \sqrt{\pi \epsilon_+^3},
\end{equation}
\begin{equation}
  \label{B2collo}
  B^{\mrm{cl}}_2 = \frac{\pi}{3} \epsilon_+^3 \Bigl\{ H\bigl(\epsilon_+ \bigr)
  - \frac{1}{2} \ln\bigl(36 \pi \epsilon_+ \bigr) \Bigr\}
  +\frac{\pi}{2} \epsilon_+,
\end{equation}
\begin{equation}
  \label{B2logcl}
  B^{\mrm{cl}}_{2\mrm{log}} = -\frac{\pi}{6} \epsilon_+^3.
\end{equation}
It is to be noted that the contributions in
$\til{f}_{\mrm{id}}$ and $B_2^{\rm HC,cl}$ contain terms that scale 
with $1/\eta$.
For dilute colloidal systems therefore the ideal contribution of the 
counterions dominates the free energy and cannot
be neglected. Effects due to the electrostatic interaction between the ions
are corrections  to the ideal behavior.

At this point we briefly introduce yet another model that is also widely used to describe
charged systems. It is the one-component plasma (OCP), which in its simplest form 
consists of a collection of $N$ equally charged point-like 
particles immersed in a neutralizing background
that ensures the global charge neutrality of the system 
(in the TCPHC electroneutrality is ensured
by oppositely charged particles). 
The OCP, or its quantum mechanical counter-part (``jellium'')
has been used in different contexts in physics, as for instance
to describe degenerate stellar matter 
(the interior of white dwarfs or the outer layer of neutron
stars) and the interior of massive planets like Jupiter. Another example comes from
condensed matter physics, where jellium is often
used as a reference state when calculating the electronic structure of 
solids. For reviews see Refs.~\cite{ishimaru:82,baus-hansen:80,abe:59}.

\begin{figure}
  \begin{center}
    \epsfig{file=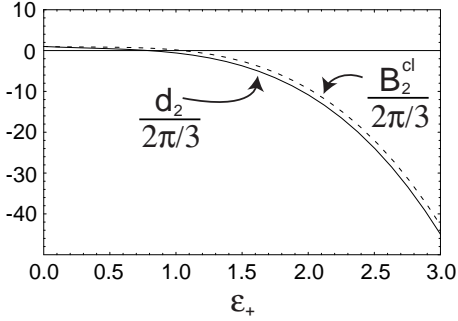, width=60mm}
  \end{center}
  \caption{Coefficients $B^{\rm cl}_2$ (Eq.~(\ref{bcocp2}))
    and $d_2$ (pure OCPHC from Ref.~\cite{roland-orland:tcp})
    as functions of the coupling $\epsilon_+$. Notice that
    both coefficients are normalized by the second virial of
    a pure one-component hard-core gas.}
  \label{fig:fig3}
\end{figure}

When the particles have a hard core, the OCP is called one-component hard-core plasma (OCPHC): what
we will see next is that the electrostatic contribution to the free energy in dilute colloidal 
suspensions can be almost described through the OCPHC.
If we compare the coefficients Eqs.~(\ref{BDHcollo}--\ref{B2log}) with the ones previously 
obtained\cite{roland-orland:tcp} for the low density expansion of the OCPHC
\begin{equation}
  \label{focpRol}
  \til{f}_{\rm OCPHC}=\til{c} \ln \til{c} -\til{c} + d_{3/2} \til{c}^{3/2} +
  d_2 \til{c}^2 + d_{\ln2} \til{c}^2 \ln \til{c} + O(\til{c}^{5/2}),
\end{equation}
where $\til{c}$ is the volume fraction of particles in the OCPHC, we find 
(cf.\ Eqs.~(51), (56) and (57) of Ref.~\cite{roland-orland:tcp})
\begin{equation}
  \label{bcocp1}
  B^{\rm cl}_{\mrm{DH}} = d_{3/2},
\end{equation}
\begin{equation}
  \label{bcocp2}
  B^{\rm cl}_2 = d_2 + \frac{\pi}{2} \epsilon_+
\end{equation}
and
\begin{equation}
  \label{bcocp3}
  B^{\rm cl}_{2\mrm{log}} = d_{\ln2},
\end{equation}
where the notation used in Ref.~\cite{roland-orland:tcp} 
is kept in Eq.~(\ref{focpRol}) and on the rhs of
Eqs.~(\ref{bcocp1}--\ref{bcocp3}). The comparison between 
$B^{\rm cl}_2$ and $d_2$
is shown in Fig.~\ref{fig:fig3}, where we rescaled both 
coefficients by their value at vanishing coupling,
$d_2(\epsilon_+ =0)=B_2^{\rm cl} (\epsilon_+ =0)= 2 \pi/3$.
This shows that, including the order $\wti{c}_+^2 \ln \wti{c}_+$ (i.e., 
for very dilute colloidal suspension), the OCPHC is almost recovered,
except an additional term in Eq.(\ref{bcocp2}).
This additional term can be quite important at intermediate values
of $\epsilon_+$ since it changes the sign of the virial coefficient,
as seen in Fig.~3.

\begin{figure}
  \begin{center}
    \epsfig{file=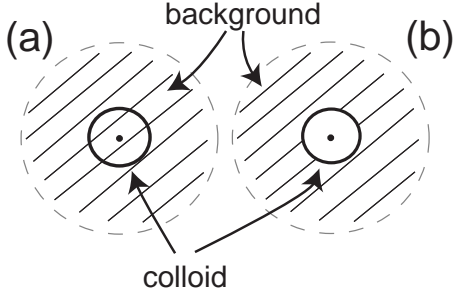,width=60mm}
  \end{center}
  \caption{A colloidal particle (with diameter $d_+$ and charge valence  $q_+$) 
in the OCPHC model
    (a) with and (b) without the penetrating background.}
  \label{fig:bolas_ocp}
\end{figure}

This is in fact what one would intuitively expect: each macroion has around it a very
large number of small neutralizing counterions which act like a background. 
Our results show that the background formed by counterions, which is a deformable
background since the counterion distribution is not uniform, acts almost like a
rigid homogeneous background, as it is assumed in OCP calculations.
However, there is 
a small difference between the TCPHC in this limit and the OCPHC: in the latter, the background
penetrates the particles, while in the TCPHC it 
cannot (for a thorough discussion of this difference see \cite{Hansen}). 
In our calculation, this is 
reflected in the $\wti{c}_+^2$ term, where $\pi \epsilon_+ / 2$ is the positive
extra cost in the free energy that the OCPHC has to pay (at this order) 
to expel the background from the hard-core particles. 
The first consequence of the non-penetrating background
 is that the effective charge of the colloids 
increases by exactly the amount of background charge that is expelled from the colloidal
interior. A simple calculation shows that the increased effective charge $q_+^{\rm eff}$ 
of colloids for the non-penetrating background
turns out to be
\begin{equation} \label{qeff}
q_+^{\rm eff} = \frac{q_+}{1- \pi \tilde{c}_+ /6}.
\end{equation}
In the low-density expansion of the OCPHC, Eq.(\ref{focpRol}),
one would have to reexpand all coefficients with
respect to the density, using  Eq.(\ref{qeff}). However, since the leading 
term which depends on the charge $q_+$ goes like $\tilde{c}_+^{3/2}$, 
the effect would come in 
at order $\tilde{c}_+^{5/2}$ and therefore is not responsible for the 
additional term in Eq.(\ref{bcocp2}).
The reason for the extra term in Eq.(\ref{bcocp2}) 
has to do with an increase of the 
self-free-energy of a colloidal particle, which can be understood in
the following way:
Assume that the OCPHC is very dilute,
such that each colloidal particle and its neutralizing background form a neutral entity 
(in the spirit of the cell model, see for instance Ref.~\cite{alexander-al:84}) 
that can be regarded independent from the other particles. 
The free energy difference \emph{per colloidal particle} between a system
without penetrating background and with penetrating background (Fig.~\ref{fig:bolas_ocp})
is then given by two contributions: one coming from the entropy lost by the 
background (formed by $N_-$ counterions) since it cannot penetrate the macroions, given by
\begin{equation}
  \frac{\Delta F_{en}}{N_+ k_B T} = 
-\frac{N_-}{N_+} \ln(1- \pi \tilde{c}_+/6) =
\frac{N_-}{N_+} \frac{\pi \til{c}_+}{6} + O(\til{c}^2_+).
\end{equation}
This is one of the contributions to 
the second virial coefficient due to the hard-core interaction, and
already included in Eq.~(\ref{f-collo-hc}). This term
does not depend on the coupling $\epsilon_+$, and therefore has nothing to do
with the extra term in Eq.(\ref{bcocp2}).
The second contribution to the self-energy is electrostatic  in origin.
Defining the charge distributions for the situations in Fig. 4
as ${\rho}_a(\vec{r})= q_+ \delta(\vec{r}) - \rho_0$ 
when the background can enter the colloids and 
${\rho}_b(\vec{r})= q_+ \delta(\vec{r}) - \rho_0 \bigl[1-\theta(d_+/2 - r)]$
when the background cannot enter the colloids
(where $\rho_0=c_+ q_+$ is the background charge density
and $\theta(x)=1$ if $x>0$ and $0$ otherwise), the electrostatic self-energy  difference 
between the two cases reads
\begin{multline}
  \frac{\Delta F_{el}}{N_+ k_B T} = \frac{1}{2} \int {\rm d}\vec{r} {\rm d}\vec{r}' \,
  \biggl\{ {\rho}_b(\vec{r}) v_c(\vec{r}-\vec{r}') {\rho}_b(\vec{r}') \\ -
  {\rho}_a(\vec{r}) v_c(\vec{r}-\vec{r}') {\rho}_a(\vec{r}') \biggr\}  \\
  = \rho_0 q_+ 4 \pi \ell_B \int_{0}^{d_+/2} {\rm d}r \, r  + O(c_+^2)  \\
  = \frac{\pi}{2} \til{c}_+ \epsilon_+ + O(c_+^2).
\end{multline}
It follows that the free energy difference per volume is given by
\begin{equation}
\Delta \tilde{f}  = \frac{ \tilde{c}_+ \Delta F_{el}}{N_+ k_BT}
= \frac{\pi \tilde{c}_+^2 \epsilon_+}{  2}. 
\end{equation}
Note that 
this is a positive contribution to the $\til{c}^2$ term
with the same coefficient as the extra term in Eq.~(\ref{bcocp2}).
Therefore, we conclude that this extra term is due to the electrostatic energy
associated with expelling the background from the colloidal volume.
This is in accord with more general results on the difference between 
OCPHC models with penetrating and excluded neutralising backgrounds\cite{Hansen}. 
In summary, whenever using the OCPHC to describe highly asymmetric charged systems
such as colloids, one has to to take into account the exclusion of the background from the
macroions. As we demonstrated, if this is taken into account, then
the TCPHC maps exactly (at least up to order $\wti{c}_+^2$) onto the OCPHC.

\subsection{Ionic activity and diameters}
\label{chapterD:radii}

In electrochemistry, it is usual to define the mean activity $\lambda_{\pm}$ 
of an electrolyte as
\begin{equation}
  \label{ma}
  \lambda_{\pm} \equiv \Bigl[ \lambda_+^{q_-} \lambda_-^{q_+} 
  \Bigr]^{ 1 / [q_+ + q_-]},
\end{equation}
where $\lambda_+$ and $\lambda_-$ are respectively the fugacities
of the positive and the negative ions. The mean activity coefficient $\mrm{f}_{\pm}$
is the ratio between the mean activity of the electrolyte and that
of an ideal gas, in general given by
\begin{equation}
  \label{mac}
  \mrm{f}_{\pm} =
  \exp \Bigl( \frac{q_-}{q_+ + q_-} 
  \frac{\partial \wti{f}_{\mrm{ex}}}{\partial \wti{c}_+} \Bigr),
\end{equation}
for a two-component system where the positively charged particles are 
used as reference species in the same way as in Eq.~(\ref{f}).
$\wti{f}_{\mrm{ex}}$, the excess free energy,
 is the difference between the free energy of the interacting systems and the free
energy of an ideal gas, i.e.,  $\wti{f}_{\mrm{ex}} = \wti{f} - \wti{f}_{\mrm{id}}$.

There are different ways of measuring $\mrm{f}_{\pm}$, as for instance, 
through the change of the 
freezing point of the solvent (usually water) 
with the addition of salt\cite{robinson-stokes:book},
by measuring the change of the potential difference 
between the electrodes of a concentration cell
as salt is added\cite{butler-roy:91,robinson-stokes:book}
(potentiometry), or by direct measurement of the solvent activity through vapor
exchange between a solution with known activity and the 
sample\cite{rard-platford:91,robinson-stokes:book} (isopiestic). 
Although dating from the early nineteen hundreds, 
these are still the most common techniques used today,
especially the potentiometry, which is regarded as the most precise technique of all. 
The values of $\mrm{f}_{\pm}$ are tabled as function of the salt concentration for many 
different electrolytes\cite{conway:book,parsons:book,robinson-stokes:book}.

From the free energy Eq.~(\ref{f}) and the definition Eq.~(\ref{mac})
we can obtain the low density expansion for the mean activity
coefficient of a $q_+ \! : \! q_-$ salt. To compare with experimental results, it is useful to
note that $\wti{c}_+ = 6.022 \times 10^{-4} q_- d_+^3 \varrho$, where $d_+$ is in {\AA}ngstr{\"o}ms
and $\varrho$ is the salt concentration in moles/liter. After the appropriate expansion,
$\mrm{f}_{\pm}$ reads
\begin{multline}
  \label{mac-expd}
  \mrm{f}_{\pm} = 1 + \nu_{\mrm{DH}} \varrho^{1/2} +
  \nu_1 \varrho +\nu_{1\mrm{log}} \varrho \ln \varrho +
  \nu_{3/2} \varrho^{3/2} \\ + 
  \nu_{3/2\mrm{log}} \varrho^{3/2} \ln \varrho + O(\varrho^2)
\end{multline}
(the order $3/2$ in the mean activity coefficient 
is the one consistent with a  free energy given up to $5/2$).
The coefficients $\nu_{\mrm{DH}}$, $\nu_1$, etc.\ are given in 
the appendix, Eqs.~(\ref{a_nuI}--\ref{a_nuE}). 
Experimentally determined activity coefficients are typically measured at 
constant pressure, while the theoretical results are obtained for constant volume.
Note that in the limit of vanishing density,
the distinction between the activities  calculated in the Lewis-Randall or in 
the McMillan-Mayer descriptions (constant pressure versus constant volume ensemble) becomes
negligible\cite{pailthorpe-al:82} in comparison to the first corrections to ideal
behavior. The experimental data at low density can be directly compared
with the theory without the need to convert between the two ensembles.

At infinite dilution, the mean activity coefficient  Eq.~(\ref{mac-expd})
goes to $1$, which is the prediction for an ideal gas. 
The first correction to the ideal
behavior is the term $\nu_{\mrm{DH}} \varrho^{1/2}$, which is the prediction one obtains from the
Debye-H{\"u}ckel limiting law (DHLL), and is independent of the ionic diameters,
see Eq.~(\ref{a_nuI}). This means that there is always a range of concentrations where
different salts (but with the same $q_+$ and $q_-$) will deviate from ideality, but have the same
activity. At higher concentrations, other terms have to be taken into account, and the ionic 
sizes begin to play an important role. In fact, as one fixes
$\ell_B$ ($\simeq 7.1$~{\AA} in water at 25~$^o$C) and the valences $q_+$ and $q_-$, 
the only free parameters in the rhs of Eq.~(\ref{mac-expd}) are
the ionic diameters $d_+$ and $d_-$, which can be used in the theoretical predictions to fit the
experimental values. 
This leads to effective equilibrium values of the ionic diameters (when in solution), 
which we call the ``thermodynamic diameter'', 
in contrast to the bare diameter\cite{israelachvili:book}
(obtained through crystallographic methods) and the hydrodynamic diameter (obtained from mobility
measurements\cite{wennerstroem:book}). By construction, it is this diameter which should be used
in equilibrium situations
if one wants to describe an electrolyte solution as a TCPHC, as for example 
computer simulations of electrolytes.

\subsubsection{Fitting assuming one mean diameter}

\begin{figure}
  \begin{center}
    \epsfig{file=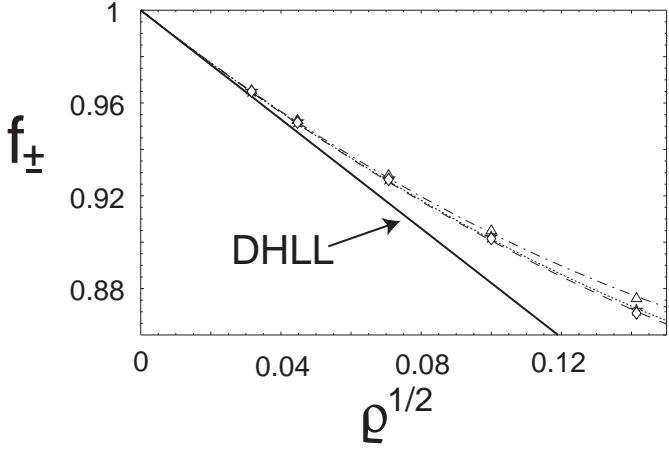, width=88mm}
  \end{center}
  \caption{Experimental\cite{sheldlovsky:50,conway:book} and 
fitted theoretical mean activity coefficient 
    $\mrm{f}_{\pm}$ for various salts as a function of the density $\varrho$ (in mole/liter),
    for various $1:1$ salts, viz.\ HCl (triangles and dot-dashed line, $d=4.0$~{\AA}),
    NaCl (stars and dotted line, $d=3.7$~{\AA}) and KCl (diamonds and dashed line, $d=3.6$~{\AA}).
    The full line denotes the Debye-H{\"u}ckel limiting law, which scales with $\varrho^{1/2}$.
    The fit was done using the data points shown (up to $\varrho = 0.02$ mole/liter).}
  \label{fig:fit1_11}
\end{figure}

We now show the fitting procedure assuming that $d_+ = d_- = d$, where $d$ is the mean diameter. 
We also restrict here this procedure to $1:1$ salts, where more experimental data is 
available at reasonably low densities (below $\sim 0.05$ mole/liter). 
For asymmetric salts, this method demands precise
data at the range below $\sim 0.01$ mole/liter, where Debye-H{\"u}ckel is often assumed 
to account for all effects and few experimental points are available.

The assumption of equal ionic sizes has been
often used in the past to fit activities to modified Debye-H{\"u}ckel
theories\cite{robinson-stokes:book}
to account for the ionic sizes (as previously mentioned, the Debye-H{\"u}ckel 
limiting law is insensitive to it).
Since we force the two diameters to be equal, we only 
need the expansion for $f_{\pm}$ up to linear order
since this will determine uniquely the single unknown parameter.
From the experimental data
$\mrm{f}^{exp}_{\pm}$\cite{sheldlovsky:50,conway:book} we
subtract the diameter-independent term and obtain the difference
\begin{equation}
    \label{delta-mac1}
    {\Delta}\mrm{f}_{\pm} \equiv \mrm{f}^{exp}_{\pm} - \nu_{\mrm{DH}} \varrho^{1/2}
    = \nu^{exp}_1 \varrho + O(\varrho^{(3/2)}).
\end{equation}
The fit to this function leads to the coefficient of the linear term $\nu^{exp}_1$,
and that can be used to determine the diameter by solving 
the equation $\nu_1 (d) = \nu^{exp}_1$ (with $d$ as unknown).
In Fig.~\ref{fig:fit1_11} we show the experimental\cite{sheldlovsky:50,conway:book} 
$f_{\pm}$ for HCl, NaCl and KCl and the theoretical results
(up to linear order in $\varrho$) using, respectively, the diameters $4.0$~{\AA}, $3.7$~{\AA} and $3.6$~{\AA}.
These values come from the fitting described above applied to the experimental data in the
range $\varrho = 0.01$ to $0.05$ mole/liter.
The diameters obtained are very close to the ones obtained 
in Ref.~\cite{sheldlovsky:50} with a similar fitting, 
but using the DH theory with an approximate way to incorporate the ionic sizes. 

The fact that $d_{\mrm{HCl}} > d_{\mrm{NaCl}} > d_{\mrm{KCl}}$, 
which is the opposite to the sequence of bare diameters, is
a consequence of the existence of hydration shells around the ions
which tend to be larger for smaller bare ion size\cite{israelachvili:book}.
Notice however that the values obtained here
lie between the bare diameters and the hydrated values available in the 
literature. This is not surprising since the effective hard-core size simultaneously
reflects both the presence of the hydration shells and the ``softness'' of the outer
water layer  as two oppositely charged ions approach each other.

\subsubsection{Fitting assuming two diameters}

If we now assume that both $d_+$ and $d_-$ are unconstrained, we have to use one more term in the
expansion of the activity coefficient and solve the coupled equations
\begin{equation}
  \label{system}
  \begin{split}
    \nu_1(d_+,d_-) &= \nu^{exp}_1 \\
    \nu_{3/2}(d_+,d_-) &= \nu^{exp}_{3/2}
  \end{split}
\end{equation}
where the coefficients $\nu^{exp}_1$ and $\nu^{exp}_{3/2}$
are obtained by fitting the function
\begin{eqnarray}
{\Delta}\mrm{f}_{\pm} &= &\mrm{f}^{exp}_{\pm} - \nu_{\mrm{DH}} \varrho^{1/2} - 
\nu_{1\mrm{log}} \varrho \ln \varrho  \nonumber \\
& =& \nu_1^{exp} \varrho + \nu^{exp}_{3/2} \varrho^{3/2} + 
O(\varrho^{3/2} \ln \varrho).
\end{eqnarray}
For the actual fitting we divide by the density and obtain
\begin{equation} \label{fitdefine}
  \frac{\Delta \mrm{f}_{\pm}}{\varrho} = \nu^{exp}_1 + \nu^{exp}_{3/2} \varrho^{1/2}.
\end{equation}
Note that we subtracted  the term proportional to $\varrho \ln \varrho$,
since like the DH term it is independent of the ionic diameters.
What is interesting about this fitting procedure is that the two ionic
diameters are independent parameters, that is, the effective sizes 
obtained through this fitting do not depend
on the size of a ``reference'' ion, contrary to what happens when using crystallographic
methods\cite{schwister-al:book}.
The method we show here can in principle lead to useful results,
as long as the experimental data is accurate enough
at very low densities, as we now demonstrate.

\begin{figure}
  \begin{center}
    \epsfig{file=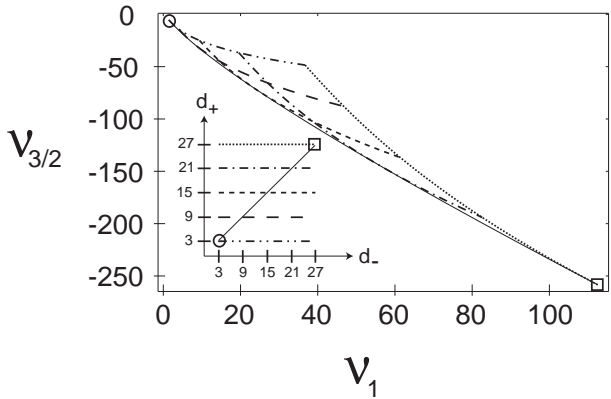, width=80mm}
  \end{center}
  \caption{Mapping ionic diameters of a $1:1$ salt in water at room temperature 
    into coefficients of the rescaled activity coefficient 
$\Delta \mrm{f}_{\pm}$ defined
in Eq.(\ref{fitdefine}).
    The inset shows different paths in the ($d_-$;$d_+$)--space 
    (with the ionic diameters between $3$~{\AA} and $27$~{\AA}) which are mapped into
    lines in the ($\nu_{1}$;$\nu_{3/2}$)--space, as determined by
 Eqs.~(\ref{a_nu1}) and (\ref{a_nu32}). By knowing the 
    values of $\nu_1$ and $\nu_{3/2}$ for a certain salt at very low densities, 
    one can make the ``inverse-mapping'' and extract the values of $d_-$ and $d_+$.
The solid line, the symmetric case with $d_+=d_-$, is the lower envelope off all
other lines which are obtained by varying one diameter while keeping the other fixed.  }
  \label{fig:map}
\end{figure}

\begin{figure}
  \begin{center}
    \epsfig{file=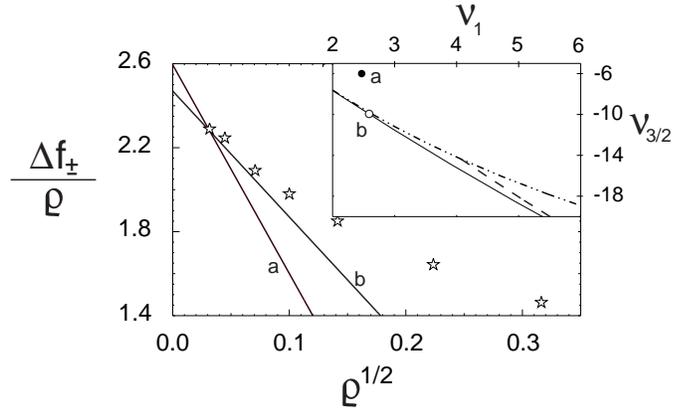, width=88mm}
  \end{center}
  \caption{Fitting procedure (NaCl in water at room temperature) 
    for two unconstrained diameters. Lines $a$ and $b$ correspond to
    two possible asymptotic fits 
    of $\Delta \mrm{f}_{\pm}/\varrho$ as $\varrho \rightarrow 0$ (see text).
    From these lines one can extract $\nu_1$ and $\nu_{3/2}$, 
which allows determination of the ionic diameters
(inset, which shows a part of the mapping done in Fig.~\ref{fig:map}) 
    Notice that while the two lines are reasonable asymptotic fits to the data, 
one (line $b$) leads
    to acceptable values for the ionic diameters, $d_+=3.8$~{\AA} and $d_-=5.4$~{\AA},
while the other (line $a$) leads to values that lie outside the range $3$ to $27$~{\AA} 
    (which is not reasonable). This means that more precise data
    in the range $\varrho < 0.01$ mole/liter is needed to obtain the correct values for the diameters
    with this method. Experimental data from Ref.~\cite{conway:book}.}
  \label{fig:mac2}
\end{figure}

Fig.~\ref{fig:map} shows 
the mapping between the parameter space of ionic diameters and  the two coefficients 
$\nu_1$ and $\nu_{3/2}$ (for a $1:1$ salt in water at room temperature).
The mapping is restricted to the range 
$3$~{\AA} $ < d_+ < 27$~{\AA} and $3$~{\AA} $ < d_- < 27$~{\AA}. This allows the ``inverse mapping''
between the ($\nu_{1}$;$\nu_{3/2}$)--space and the ($d_-$;$d_+$)--space: any system
with ionic diameters within the values above should have the coefficients $\nu_1$ and $\nu_{3/2}$
inside the region in the ($\nu_{1}$;$\nu_{3/2}$)--space, as shown in Fig.~\ref{fig:map}.
One problem becomes obvious: the coefficient $\nu_{3/2}$ is larger than the 
coefficient of the linear term, $\nu_1$, meaning that experimental data at very low 
concentrations are needed.
As an example, Fig.~\ref{fig:mac2} shows $\Delta \mrm{f}_{\pm} / \varrho$ 
for NaCl\cite{conway:book} as a function of $\varrho^{1/2}$, which should be
asymptotically linear in the limit  $\varrho \rightarrow 0$ and therefore give the coefficients
$\nu_1$ and $\nu_{3/2}$ by a simple linear fit according to Eq.(\ref{fitdefine}).
The lines shown are two possible asymptotic linear fits
to  the experimental points (line $a$ is given by 
$\Delta f_{\pm} / \varrho = 2.59-9.94 \varrho^{1/2}$, and line $b$ by
$ 2.47-6.00 \varrho^{1/2}$). The inset to Fig.~\ref{fig:mac2} gives the corresponding
positions in the ($\nu_{1}$;$\nu_{3/2}$)--space (cf.\ Fig.\ref{fig:map})
for the two lines. Clearly, only one of the fits (line $b$) leads to reasonable
values for the diameters (between $3$ and $9$~{\AA} 
as seen from the position of point $b$ in the inset,
$d_+=3.8$~{\AA} and $d_-=5.4$~{\AA} as obtained by solving Eq.~(\ref{system})), while the
other fit (line $a$) leads to unreasonable values for the diameters (outside the range
$3$ to $27$~{\AA}). In other words, the asymptotic extrapolation of the experimental data
is very sensitive to small errors, and in order to obtain the diameters with this method
one needs more accurate data at very low densities 
(which to the best of our knowledge is not available in the
literature). Although NaCl was used as example, the situation is identical for other salts.

%%%%%%%%%%%%%%%%%%%%%%%%%%%%%%%%%%%%%%%%%%%%%%%%%%

\section{Conclusions}
\label{chapterD:conclusions}

Using field theoretic methods we obtained the exact low density (``virial'')
expansion of the TCPHC up to order $5/2$ in density. In its general form, the model can be applied
to both electrolyte solutions and dilute colloidal suspensions (when the van der Waals forces are
unimportant); the free energy derived here provides a
 unified way for handling both systems in the limit of
low concentration. As the calculations show, the generalization to short-ranged potentials
other than hard core is possible.

The behavior of the coefficients $B_2$ and $B_{5/2}$ suggests that the series is badly convergent,
meaning that the inclusion of higher order terms does not necessarily extend the validity of the
free energy to larger densities. We saw that the divergent behavior of these coefficients is
related to the ionic pairing\cite{bjerrum:26,fisher-levin:93,yeh-zhou-stell:96} which is favored
as the coupling increases (this is not present e.g.\ in the OCPHC\cite{roland-orland:tcp}).
This also means that 
such a low-density expansion is not very useful to study the phase 
behavior\cite{three-component} of ionic systems: 
in the situation of phase-separation between a 
very dilute phase and a dense phase, typically 
the average density in the dense phase already falls
outside the range of validity of the low density expansion.

In applying our results to colloidal systems we concluded 
that at low density the counterion
entropic contribution dominates over the electrostatic 
contribution due to the macroions; the latter
contribution can be described, in this limit and up to this order, 
by an OCPHC corrected by effects due to the exclusion of
counterions from the colloidal particles.

Finally, we used the theoretical results for the 
mean activity coefficient to fit experimental data
and extract effective ionic sizes. In the simplest fitting, 
where we assumed the two ionic diameters
to be equal, we obtained sizes that are reasonable and close to what 
one would expect from the results
obtained by other methods. For the more interesting case 
where the two ionic sizes are taken as free
parameters and determined independently, we showed that one would 
need more experimental data 
for  the mean activity coefficient at very low densities (which, to the best of our knowledge, 
is not available in the literature) to obtain the correct values for the diameters. With the proper
experimental data it would then be a simple matter to obtain the effective thermodynamic
ionic sizes, which could serve as useful input for computer simulations of 
two-component-plasmas with hard-cores to model electrolyte solutions.

%%%%%%%%%%%%%%%%%%%%%%%%%%%%%%%%%%%%%%%%%%%%%%%%%%

\begin{acknowledgement}
We thank N.\ Brilliantov and H.\ L{\"o}wen for useful discussions.
AGM acknowledges the support from FCT through grant PRA\-XIS XXI/BD/13347/97 and
DFG.
\end{acknowledgement}

%%%%%%%%%%%%%%%%%%%%%%%%%%%%%%%%%%%%%%%%%%%%%%%%%%

\section*{Appendix}

\subsection*{Averages needed for $Z_1$ and $Z_2$}

The following expressions were used to obtain Eqs.~(\ref{Z1}--\ref{Zlast}).
In order to have more compact formulas we use
the Greek letters $\alpha$ and $\beta$ instead of $+$ or $-$. For instance,
$\alpha \, q_{\alpha}$ means both $+ q_+$  and $- q_-$.
\begin{equation}
  \label{a_av_1}
  \aveb{h_{\alpha}(\vv{r})} = \ee{ q_{\alpha}^2 v_c(0) / 2 },
\end{equation}
\begin{equation}
  \aveb{\ee{-\imath \alpha q_\alpha \phi(\vv{r})}} = 
  \ee{-q_{\alpha}^2 v_{\mrm{DH}}(0) / 2},
\end{equation}
\begin{equation}
  \aveb{h_{\alpha}(\vv{r}) h_{\beta}(\vv{r}')} = 
  \ee{- \omega_{\alpha \beta}(\vv{r}-\vv{r}') + 
    \bigl[ q_{\alpha}^2 v_c(0) + q_{\beta}^2 v_c(0) \bigr] / 2} ,
\end{equation}
\begin{equation}
  \begin{split}
    \aveb{\ee{-\imath \alpha \, q_{\alpha} \phi(\vv{r}) - \imath \beta \,
        q_{\beta} \phi(\vv{r}')}} &= 
    \ee{- \bigl[ q_{\alpha}^2 v_{\mrm{DH}}(0) + q_{\beta}^2 v_{\mrm{DH}}(0) \bigr] / 2} \\ & \times
    \ee{-\alpha  \beta \, q_{\alpha} q_{\beta} v_{\mrm{DH}}(\vv{r}-\vv{r}')}
  \end{split}
\end{equation}
\begin{equation}
  \label{a_av_last}
  \aveb{\phi^2(\vv{r}') \ee{- \imath \alpha \, q_{\alpha} \phi(\vv{r})} } =
  \ee{-q_{\alpha}^2 v_{\mrm{DH}}(0) / 2} \Bigl[ v_{\mrm{DH}}(0) -
  q_{\alpha}^2 v^2_{\mrm{DH}}(\vv{r}-\vv{r}') \Bigr].
\end{equation}

The brackets $\langle \cdots \rangle$ denote averages over the 
fields $\phi$ and $\psi_\alpha$ where
$\omega^{-1}$ and $v_{\mrm{DH}}^{-1}$ are the propagators.

\subsection*{The coefficients in the grand-canonical free energy}

We give here the explicit expressions for the coefficients 
$\tilde{g}$, Eq.~(\ref{free-energy3}).
Using the function $H(x)$, defined in Eq.(\ref{Hdefine}),
the coefficients read
\begin{equation}
  \label{a_coeff_1}
  m_1 = \frac{\pi}{3} \frac{q_+^6 \ell_B^3}{d_+^3}
  \biggl\{ - H\Bigl( \frac{q_+^2 \ell_B}{d_+} \Bigr) + 
  \ln \Bigl(\frac{3 d_+}{\ell_B} \Bigr) \biggr\},
\end{equation}
\begin{equation}
  m_2 = \frac{\pi}{3} \frac{q_-^6 \ell_B^3}{d_+^3}
  \biggl\{ - H\Bigl( \frac{q_-^2 \ell_B}{d_-} \Bigr)
  + \ln \Bigl(\frac{3 d_-}{\ell_B} \Bigr) \biggr\}, 
\end{equation}
\begin{equation}
  m_3 = \frac{2 \pi}{3} \frac{q_+^3 q_-^3 \ell_B^3}{d_+^3}
  \biggl\{ H\Bigl( \frac{2 q_+ q_- \ell_B}{d_+ + d_-} \Bigr)
  - \ln \Bigl(\frac{3 [d_+ + d_-]}{2 \ell_B} \Bigr) \biggr\},
\end{equation}
\begin{equation}
  n_1 = \frac{\pi}{3} \frac{q_+^8 \ell_B^3}{d_+^3}
  \biggl\{ - \frac{5}{8} - 2 H\Bigl( \frac{q_+^2 \ell_B}{d_+} \Bigr)
  + \ln \Bigl(\frac{12 d_+^2}{\ell_B^2} \Bigr) \biggr\},
\end{equation}
\begin{equation}
  n_2 = \frac{\pi}{3} \frac{q_-^8 \ell_B^3}{d_+^3}
  \biggl\{ - \frac{5}{8} - 2 H\Bigl( \frac{q_-^2 \ell_B}{d_-} \Bigr)
  + \ln \Bigl(\frac{12 d_-^2}{\ell_B^2} \Bigr) \biggr\}, 
\end{equation}
\begin{equation}
  \begin{split}
    n_3 &= \frac{2 \pi}{3} \frac{q_+^3 q_-^3 \ell_B^3}{d_+^3}
    \biggl\{ - \frac{5}{8} q_+ q_- + \frac{[q_+ - q_-]^2}{2} 
    H\Bigl( - \frac{2 q_+ q_- \ell_B}{d_+ + d_-} \Bigr)  \\
    &+ q_+ q_- \ln \Bigl(\frac{2 [d_+ + d_-]}{\ell_B} \Bigr) -
    \frac{q_+^2 + q_-^2}{2} \ln \Bigl(\frac{3 [d_+ + d_-]}{2 \ell_B} \Bigr) \biggr\}, 
  \end{split}
\end{equation}
\begin{equation}
  p_1 = \frac{\pi}{3} \frac{q_+^6 \ell_B^3}{d_+^3},
  \qquad
  p_2 = \frac{\pi}{3} \frac{q_-^6 \ell_B^3}{d_+^3},
  \qquad
  p_3 = - \frac{2 \pi}{3} \frac{q_+^3 q_-^3 \ell_B^3}{d_+^3},
\end{equation}
\begin{equation}
  r_1 = \frac{2 \pi}{3} \frac{q_+^8 \ell_B^3}{d_+^3},
  \qquad
  r_2 = \frac{2 \pi}{3} \frac{q_-^8 \ell_B^3}{d_+^3},
\end{equation}
\begin{equation}
  r_3 = \frac{2 \pi}{3} \frac{q_+^3 q_-^3 \ell_B^3}{d_+^3} 
  \Bigl[ q_+ q_- - \frac{q_+^2 + q_-^2}{2} \Bigr],
\end{equation}
\begin{equation}
  \label{a_coeff_last}
  s_1 = \frac{q_+^4}{8},
  \quad
  s_2 = \frac{q_-^4}{8},
  \quad
  t_0 = \frac{d_+^3}{12 \pi \ell_B^3},
  \quad
  t_1 = \frac{q_+^6}{48},
  \quad
  t_2 = \frac{q_-^6}{48}.
\end{equation}

\subsection*{The coefficients in the fugacity}

In Eq.~(\ref{lambda-}) we define the fugacity of negative ions
$\wti{\lambda}_-$ as an expansion in terms of $\wti{\lambda}_+$. The coefficients
$a_0$, $a_1$, etc. are determined by the condition of global electroneutrality. 
They are given by
\begin{equation}
  \label{a_a0}
  a_0=\frac{q_+}{q_-},
\end{equation}
\begin{equation}
  a_1=\frac{q_+^2}{q_-} \bigl[ q_+^2 - q_-^2 \bigr]
  \sqrt{\frac{\pi \ell_B^3}{d_+^3} \Bigl[1 + \frac{q_-}{q_+} \Bigr]},
\end{equation}
\begin{multline}
  a_2= \frac{q_+}{q_-^2} \bigl[ -2 m_2 q_+ + 2 m_1 q_- + m_3 [q_1-q_2] \bigr] +
  \frac{\pi}{6} \frac{q_+^2 \ell_B^3}{q_- d_+^3} \bigl[q_+ + q_- \bigr]^2 \\ \times
  \bigl[ 7 q_+^3 - 9 q_+^2 q_- + 2 q_-^3 \bigr] +
  \frac{\pi}{3} \frac{q_+^2 \ell_B^3}{q_- d_+^3}   
  \bigl[ q_+^5 - q_+^3 q_-^2 + q_+^2 q_-^3
  -q_-^5 \bigr] \\ \times
  \ln \Bigl( \frac{4 \pi \ell_B^3}{d_+^3} 
  q_+ \bigl[ q_+ + q_- \bigr] \Bigr),
\end{multline}
\vspace{-5mm}
\begin{equation}
  a_3 = \frac{\pi}{3} \frac{q_+^2 \ell_B^3}{q_- d_+^3}   
  \bigl[ q_+^5 - q_+^3 q_-^2 + q_+^2 q_-^3
  -q_-^5 \bigr],
\end{equation}
\vspace{-5mm}
\begin{multline}
  a_4 = \frac{\ell_B^{3/2}}{24 q_-^3} \sqrt{\pi q_+ \bigl(q_+ + q_- \bigr)} \\ \times
  \Bigl\{ 24 n_1 \bigl[ 5 q_+ q_-^2 - q_-^3 \bigr] 
  +24 n_2 \bigl[ q_+^3 - 5 q_+ q_-^2 \bigr]
  \\ +72 n_3 \bigl[ q_+^2 q_- -  q_+ q_-^2 \bigr] 
  +24 m_1 \bigl[ q_+^2 q_-^3 - 3 q_+ q_-^4 \bigr]
  +24 m_2 \bigl[ -4 q_+^4 q_- \\ - q_+^3 q_-^2 + 7 q_+^2 q_-^3 \bigr]
  +12 m_3 \bigl[ 2 q_+^4 q_- - q_+^3 q_-^2 - 6 q_+^2 q_-^3 +5 q_+ q_-^4 \bigr] \\
  +\frac{\pi \ell_B^3}{d_+^3} q_+^2 q_-^2 [q_+ + q_-]^2
  \bigl[ 26 q_+^5 - 34 q_+^4 q_- - 31 q_+^3 q_-^2 + 67 q_+^2 q_-^3 \\ -
  45 q_+ q_-^4 + 17 q_-^5 \bigr] \Bigr\} +
  \frac{\pi^{3/2}}{6} \frac{q_+^{5/2} \ell_B^{9/2}}{q_- d_+^{9/2}}
  \bigl[q_+ - q_- \bigr] \bigl[q_+ + q_- \bigr]^{5/2} \\ \times
  \bigl[ 10 q_+^4 -11 q_+^3 q_- + 13 q_+^2 q_-^2  -4 q_+ q_-^3 + 
  3 q_-^4 \bigl] \\ \times \ln \Bigl( \frac{4 \pi \ell_B^3}{d_+^3} q_+ \bigl[q_+ + 
  q_- \bigr] \Bigr),
\end{multline}
\vspace{-6mm}
\begin{multline}
  \label{a_a5}
  a_5 = \frac{\pi^{3/2}}{6} \frac{q_+^{5/2} \ell_B^{9/2}}{q_- d_+^{9/2}}
  \bigl[q_+ - q_- \bigr] \bigl[q_+ + q_- \bigr]^{5/2}
  \bigl[ 10 q_+^4 -11 q_+^3 q_- \\ + 13 q_+^2 q_-^2 -4 q_+ q_-^3 + 
  3 q_-^4 \bigl] \ln \Bigl( \frac{4 \pi \ell_B^3}{d_+^3} q_+ \bigl[q_+ + 
  q_- \bigr] \Bigr).
\end{multline}
Notice that $m_1$, $m_2$, etc.\ were defined in 
Eqs.~(\ref{a_coeff_1}--\ref{a_coeff_last}).

\subsection*{The coefficients in the mean activity coefficient}

In Eq.~\ref{mac-expd} we obtained the low-density expansion of the
mean activity coefficient of ionic solutions where the ions have
valences $q_+$ and $q_-$ and effective diameters $d_+$ and $d_-$ (in {\AA}ngstr{\"o}ms). 
Defining $\omega \equiv 6.022 \times 10^{-4} \ell_B^3 q_+^6 q_-$
(with $\eta \equiv q_-/q_+$ and $\xi \equiv d_-/d_+$),
the coefficients in  Eq.~\ref{mac-expd} read
\begin{equation}
  \label{a_nuI}
  \nu_{\mrm{DH}} = - \eta \sqrt{\pi \omega [1+\eta]},
\end{equation}
\begin{multline}
  \label{a_nu1}
  \nu_1 = \frac{\pi \omega \eta}{6 [1+\eta]}
  \Bigl\{ -1 + 3 \eta + 8 \eta^2 + 3 \eta^3 -\eta^4 
  + 4 \Bigl[ 
  H\bigl(\epsilon_+ \bigr) +  \ln\bigl( \epsilon_+ \bigr) 
  \Bigr] \\
  +4 \eta^4 \Bigl[ 
  H\bigl(\frac{\eta^2 \epsilon_+}{\xi} \bigr) +
  \ln\bigl(\frac{\eta^2 \epsilon_+}{\xi} \bigr) 
  \Bigr]
  -8 \eta^2 \Bigl[
  H\bigl(- \frac{2 \eta \epsilon_+}{1+\xi} \bigr) +
  \ln\bigl(\frac{2 \eta \epsilon_+}{1+\xi} \bigr) 
  \Bigr] \\
  +8 \eta^2 [1-\eta^2] \ln(\eta) 
  - \bigl[ 2 - 4 \eta^2 +2 \eta^4 \bigr] 
  \ln \bigl( 36 \pi \omega [1+\eta] \bigr) \Bigl\},
\end{multline}
\vspace{-5mm}
\begin{equation}
  \nu_{1\mrm{log}}= - \frac{\pi \omega \eta}{3} [1-\eta]^2 [1+\eta],
\end{equation}
\vspace{-5mm}
\begin{multline}
  \label{a_nu32}
  \nu_{3/2} = \frac{\eta [\pi \omega]^{3/2}}{24 \sqrt{1+\eta}}
  \Bigl\{ 42 - 6\eta -14 \eta^2 +68 \eta^3 -14 \eta^4 -6 \eta^5
  +42 \eta^6 \\
  + 40 \bigl[ 1- \frac{2 \eta}{5} \bigr] \Bigl[ 
  H\bigl(\epsilon_+ \bigr) +  \ln\bigl( \epsilon_+ \bigr) 
  \Bigr]
  + 40 \eta^6 \bigl[ 1- \frac{2}{5 \eta} \bigr] \Bigl[ 
  H\bigl(\frac{\eta^2 \epsilon_+}{\xi} \bigr) \\ +
  \ln\bigl(\frac{\eta^2 \epsilon_+}{\xi} \bigr) 
  \Bigr]
  + 112 \eta^3 \Bigl[
  H\bigl(- \frac{2 \eta \epsilon_+}{1+\xi} \bigr) +
  \ln\bigl(\frac{2 \eta \epsilon_+}{1+\xi} \bigr) 
  \Bigr] \\
  - 16 \eta^3 \bigl[7 -2 \eta^2 + 5 \eta^3 \bigr] \ln(\eta)
  + 8 \eta \bigl[1-\eta^2 \bigr]^2 \ln \bigl(36 \pi \omega [1+\eta] \bigr) \\
  -20 \bigl[1+\eta^3 \bigr]^2 \ln \bigl(64 \pi \omega [1+\eta] \bigr)
  \Bigl\},
\end{multline}
\vspace{-5mm}
\begin{multline}
  \label{a_nuE}
  \nu_{3/2\mrm{log}} = - \frac{\eta [\pi \omega]^{3/2}}{6} \sqrt{1+\eta} \\ \times
  \bigl[ 5 - 7 \eta + 7 \eta^2 + 7 \eta^3 -7 \eta^4 + 5 \eta^5 \bigr].
\end{multline}

\bibliographystyle{}

\begin{thebibliography}{10}
\bibitem{note1}
This can be checked in our calculations by turning-off the charges (or simply
  be setting $\ell_B =0$) in the final result.
\bibitem{note_volfrac}
The definition of volume fraction usually found in the literature---normally
  represented as $\eta$---relates to our definition through $\eta = \pi \wti{c}
  / 6$.
\end{thebibliography}

\end{document}